\documentclass[a4paper,11pt]{article}
\pdfoutput=1 

\usepackage{jinstpub} 

\usepackage{color}
\usepackage[normalem]{ulem}
\usepackage{lineno}
\usepackage{siunitx}
\usepackage{newtxmath,newtxtext}

\newlength{\figwidth}
\newlength{\fighalfwidth}
\title{Data Unfolding with Wiener-SVD Method}

\author[a,1]{W. Tang,\note{Co-first author}}
\author[b,1]{X. Li,}
\author[a,2]{X. Qian,\note{Corresponding author}}
\author[a]{H. Wei,}
\author[a]{C. Zhang,}
\affiliation[a]{Physics Department, Brookhaven National Laboratory, Upton, NY, USA}
\affiliation[b]{State University of New York at Stony Brook, Department of Physics
  and Astronomy, Stony Brook, NY, USA}
\emailAdd{xqian@bnl.gov}

\abstract{Data unfolding is a common analysis technique used in HEP data 
analysis. Inspired by the deconvolution technique in the digital signal 
processing, a new unfolding technique based on the SVD
technique and the well-known Wiener filter is introduced.
The Wiener-SVD unfolding approach achieves the 
unfolding by maximizing the signal to noise ratios in the effective frequency
domain given expectations of signal and noise and is free from regularization parameter.
Through a couple examples,
the pros and cons of the Wiener-SVD approach as well as the nature of the
unfolded results are discussed. }


\begin{document}
\maketitle
\flushbottom

\section{Introduction}\label{sec:introduction}

Data unfolding is a common technique used in the analysis of high energy 
physics (HEP) experimental data. Some of the recent examples in the field of
neutrino physics can be found in Refs.~\cite{An:2015nua,DeVan:2016rkm,Abe:2016tmq}
and some reviews on this topic can be found in
Refs.~\cite{cowan_review,Volker_2011,Spano:2013nca,Kuusela_review}.
The motivation for
data unfolding is to estimate the true signal (e.g. energy spectrum) given
a measurement that is affected by the detector response as well as statistical (e.g.
associated with signal and backgrounds) and systematic uncertainties (e.g. associated
with backgrounds, mis-modeling of detector response due to imperfect calibration or
finite statistics in simulations).
In many applications, data
unfolding is not necessarily required. For example, in the case of a
hypothesis testing problem, it is generally more
advantageous to fold the detector response with the hypothesis and compare
with the measurement (See Ref.~\cite{Cousins:2016ksu} for more
discussions). On the other hand, the data unfolding technique is
helpful in many occasions where additional actions are required on the unfolded
results. For example, unfolded results are convenient to compare results
from different experiments that have different detector responses. Another
example would be to extract the ratio of unfolded results (such as cross
sections on different nuclei with different detector responses) to be compared
with theoretical calculations of ratios, which are typically more precise than
the calculation of individual quantities. Finally, the usage of unfolded results,
which is generally closer to the true signal than the measurement, has
natural advantages for the presentation purpose.

As explained by numerous reviews~\cite{cowan_review,Volker_2011,Spano:2013nca,Kuusela_review},
the main challenge to be overcome in the unfolding of data is the presence
of both detector smearing and uncertainties. The random fluctuations
due to the existence of statistical and systematic uncertainties
would be significantly amplified
by a naive inverse of the detector response matrix, which usually leads
to meaningless results. This is easy to understand, as the detector
smearing represents a loss of information, which in principle cannot
be recovered. In HEP, there are two main
data unfolding approaches to mitigate this issue. The first method is  the
Tikhonov regularization (or SVD unfolding)~\cite{Hocker:1995kb,Schmitt:2012kp}.
In this approach, the unfolding problem is expressed as a minimization of a
chi-square function comparing the measurement with the prediction. The large fluctuations
(also called variance) in the unfolded results are regularized by adding a penalty term into
the chi-square
function. The penalty term can be chosen to regularize the strength
or the curvature (second derivative) of the unfolded results, among other
possible choices. A parameter commonly known as the regularization
strength can be adjusted freely to control the relative size of the penalty term.
A scan of
the regularization strength is typically required to obtain the optimum value according to
certain pre-chosen metric. A common metric is the summation of variance and
bias of the unfolded results. Other choices of the metric can be found in
Ref.~\cite{cowan_review}. The second method is the expectation-maximization
iteration with early stopping (or Bayesian unfolding)~\cite{DAgostini:1994fjx}.
In this approach, one would start from an initial guess of the true signal.
During each iteration, the guess would be modified according to the difference
between the measurement and prediction given the previous guess.
Given an initial guess which is non-negative, the solution after an infinite number
of iterations approaches the result of minimizing the chi-square under positivity
constraints, which require all the unfolded data to be non-negative. This would again suffer
from large fluctuations. To mitigate that, the regularization is achieved by stopping
the iteration early before convergence. Typically, the number of iterations needs to be
scanned to achieve an optimal result. Therefore, an important issue in unfolding is to
find an appropriate trade-off between bias and variance of the estimators. 

In both unfolding approaches, a scan of the corresponding regularization parameter is required.
Inspired by the deconvolution techniques in the digital signal processing,
we propose a new unfolding method based on the Wiener filter and the SVD unfolding,
which takes into account both the expectation of signal and noise through
maximizing the signal to noise ratios in the effective frequency domain with an orthogonal
basis and avoids the scanning of any regularization parameter.  In Sec.~\ref{sec:wiener},
we review the Wiener filter in the digital signal processing employed in Liquid Argon Time
Projection Chamber (LArTPC) detectors. We then present the actual Wiener-SVD unfolding
algorithm in Sec.~\ref{sec:algorithm}. In Sec.~\ref{sec:xs} and Sec.~\ref{sec:reactor}, we
illustrate the performance of the Wiener-SVD unfolding and compare it with the
(Tikhonov) regularization method through two physics examples. The findings are
summarized in Sec.~\ref{sec:discussion}.

\section{Wiener Filter in Digital Signal Processing}\label{sec:wiener}

The problem of data unfolding shares many common features with the digital
signal processing problem, as the goal of both is to extract an estimation of
signal from the data. 
For example, in a LArTPC, the deconvolution
technique is used to ``remove'' the impact of field and electronics response from
the measured time-series signal to recover the true signal (the time profile of the
number of ionized electrons)~\cite{Baller:2017ugz,signal_processing}.
In the following, we briefly review the deconvolution technique.

Deconvolution is a mathematical technique to extract a \textit{real signal}
$S(t)$ from a \textit{measured signal} $M(t')$.  The measured signal is
modeled as a convolution integral over the real signal $S(t)$ and a
given detector \textit{response function} $R(t,t')$ which gives the
instantaneous portion of the measured signal at some time $t'$ due to
an element of real signal at time $t$ in addition to noises $N(t')$:
\begin{equation}\label{eq:decon_1}
M(t') = \int_{-\infty}^{\infty}  R(t,t') \cdot S(t) \cdot dt + N(t').
\end{equation}
If the detector response function only depends on the relative time 
difference between $t$ and $t'$,
\begin{equation}~\label{eq:sym_condition}
R(t,t') \equiv R(t'-t),
\end{equation}
 we can solve the above equation by 
doing a Fourier transformation on both sides of the equation:
\begin{equation}\label{eq:decon_2p}
M(\omega) = R(\omega) \cdot S(\omega) + N(\omega), 
\end{equation}
where $\omega$ is the frequency and $\hat{S}$ is an estimation of $S$.
We can derive the signal in the 
frequency domain by taking the ratio of the measured signal and the response function:
\begin{equation}\label{eq:decon_2}
\hat{S}(\omega) = \frac{M(\omega)}{R(\omega)} = S(\omega) + \frac{N(\omega)}{R(\omega)}.
\end{equation}
When the noise can be ignored, the real signal in the time domain can then be
obtained by applying an inverse Fourier transformation to both sides of Eq.~\ref{eq:decon_2}.

\begin{figure}[!h!tbp]
\includegraphics[width=\figwidth]{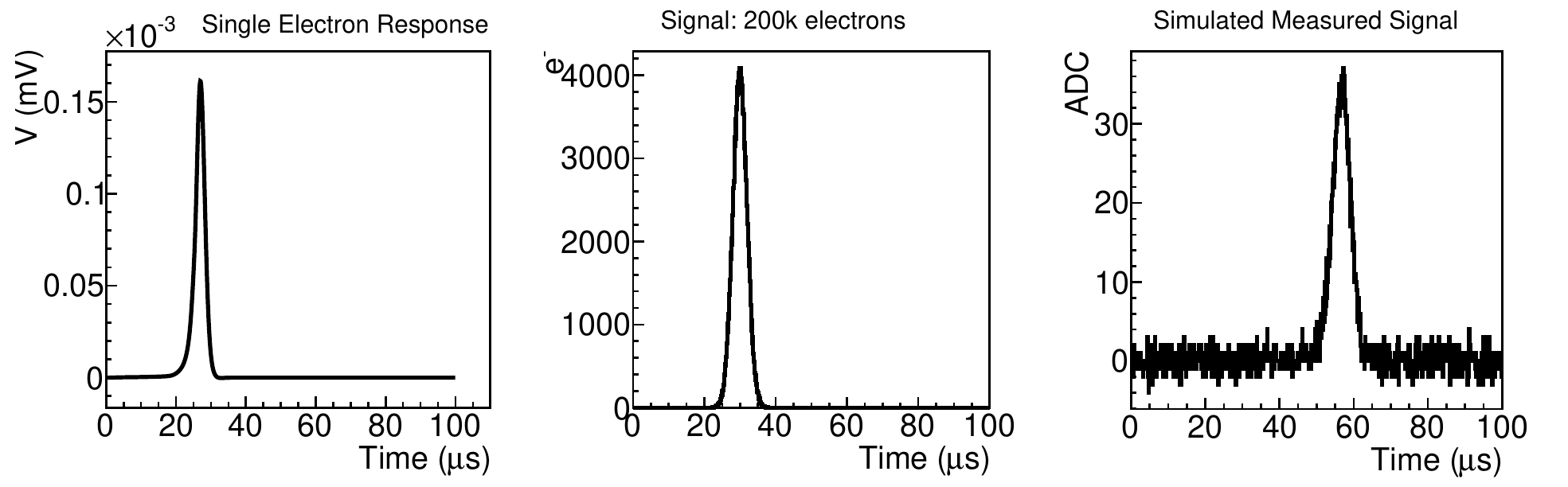}
\caption{(Left) Total detector response for a single electron is shown. (Middle)
200k electrons with a spread of 2 $\mu s$ is assumed to be the signal.
(Right) The simulated signal with electronics noise added. The electronics noise
is assumed to be white with the root-mean-square (RMS) being taken as 1.5 ADC.}
\label{fig:simu_signal}
\end{figure}

When the noise term $N(\omega)$ cannot be neglected, since the response function $R(\omega)$
is typically small at high frequencies due to the shaping of electronics, the noise components
in those frequencies will be significantly amplified by the deconvolution $N(\omega)/R(\omega)$
leading to large fluctuations in the deconvoluted signal.
Figure~\ref{fig:simu_signal} shows an example of detector response $R(t'-t)$
(left panel), true signal $S(t)$ (middle panel), and simulated measured
signal $M(t)$ with noise (right panel).

To address the issue of noise, a \textit{filter function} $F(\omega)$ is
introduced to obtain the estimator of the true signal~\cite{Baller:2017ugz,signal_processing}:
\begin{equation}\label{eq:decon_filt}
\hat{S}(\omega) = \frac{M(\omega)}{R(\omega)} \cdot F(\omega).
\end{equation}
Its purpose is to attenuate the problematic noise in the deconvolution.
The addition of this function can be considered as an augmentation to the
response function. A common choice of the filter function is the
Wiener filter~\cite{wiener}, which is constructed using the expected 
measured signal $\overline{R^2(\omega) \cdot S^2(\omega)}:=E[R^2(\omega) \cdot S^2(\omega)]$ and noise
$\overline{N^2(\omega)}:=E[N^2(\omega)]$ in the frequency domain:
\begin{equation}\label{eq:wiener}
  F(\omega) = \frac{\overline{R^2(\omega) \cdot S^2(\omega)}}{\overline{R^2(\omega) \cdot S^2(\omega)} + \overline{N^2(\omega)}},
\end{equation}
with $E\left[ \cdot \right]$ denotes the expectation operator, which can be understood
as the average after a large amount of measurements for the quantity of interest.
Note, the expectations of signal and noise squared can be viewed as prior information (e.g.
previous measurements). In practice, as we will describe in Sec.~\ref{sec:algorithm}
(Eq.~\ref{eq:signal_exp1}), one can also estimate the expectations based on the
current measurement.  The functional form of Wiener filter in Eq.~\ref{eq:wiener} is obtained by
minimizing the residual~\cite{wiener,kolmogorov}
\begin{eqnarray}
 MSE &=& E\left[ \left(F(\omega)\cdot  M(\omega) - R(\omega) \cdot S(\omega)\right)^2 \right] \nonumber \\
 &=& E\left[ \left(F(\omega)\cdot \left( R(\omega) \cdot S(\omega) + N(\omega)\right)- R(\omega) \cdot S(\omega)\right)^2 \right] \nonumber \\
 &=& F^2 \cdot \left( E\left[R^2 S^2\right] + E\left[ N^2\right]\right) + E\left[R^2 S^2 \right] - 2 F \cdot E\left[R^2 S^2\right].
\end{eqnarray}
The last step omits the $\omega$ and is obtained with the fact $E\left[N\right]\equiv 0$.
The minimization of $MSE$ (mean squared error) is achieved through $\frac{\partial MSE}{\partial F} = 0$. It is
easy to see Eq.~\ref{eq:wiener} is recovered.

\begin{figure}[!h!tbp]
\includegraphics[width=\figwidth]{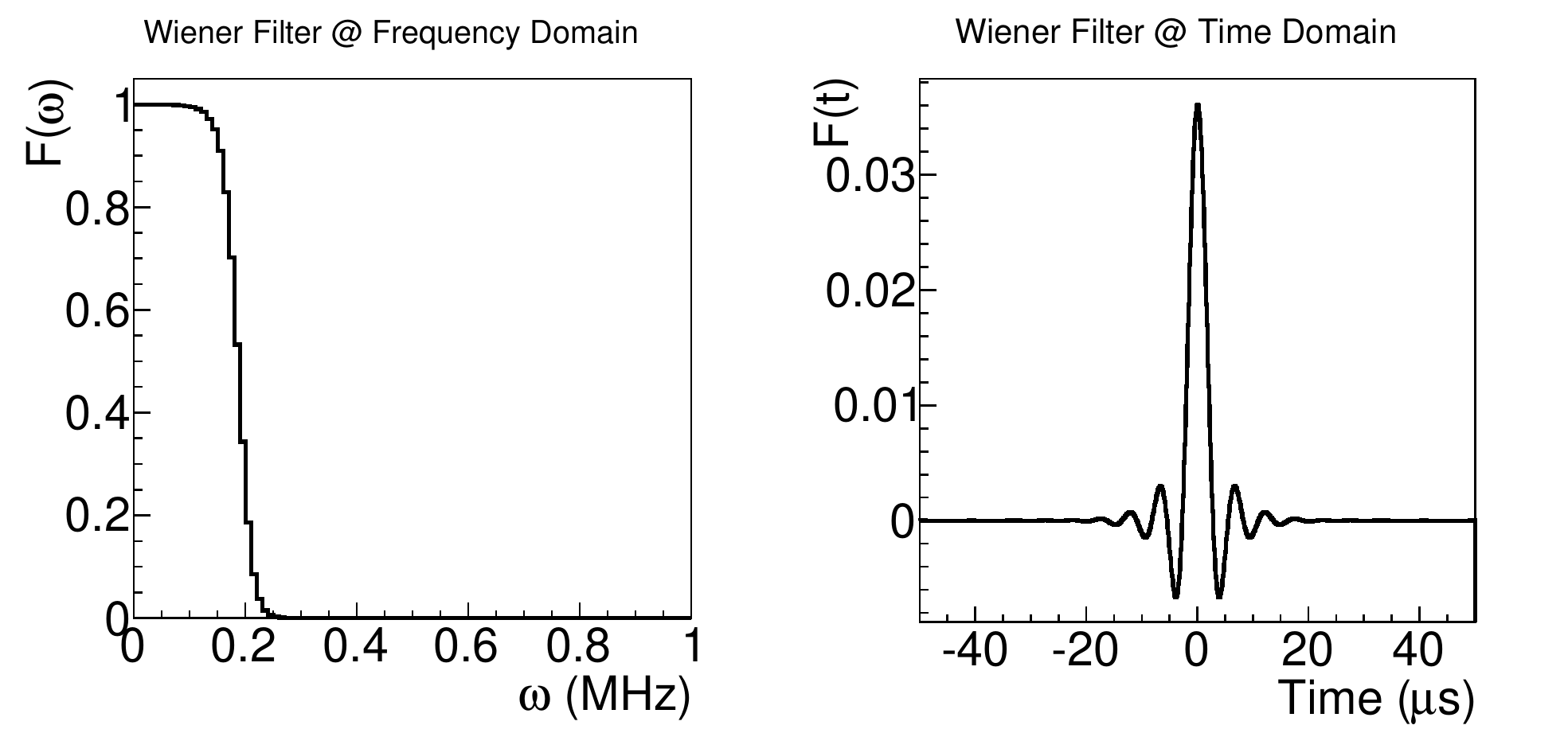}
\caption{Constructed Wiener filter in the frequency (left) and time (right) 
domains given the example shown in Fig.~\ref{fig:simu_signal}.}
\label{fig:wiener}
\end{figure}

With the construction in Eq.~\ref{eq:wiener}, the Wiener filter is expected
to achieve the best signal to noise ratio. Besides digital signal
processing, the Wiener filter is also widely used in other fields. For example, 
Wiener filters have been used in experimental astrophysics~\cite{Hu:2002qc,Simon:2009me}. Figure~\ref{fig:wiener} shows the
constructed Wiener filter in both the frequency and time domains given the
example shown in Fig.~\ref{fig:simu_signal}.

\begin{figure}[!h!tbp]
\includegraphics[width=\figwidth]{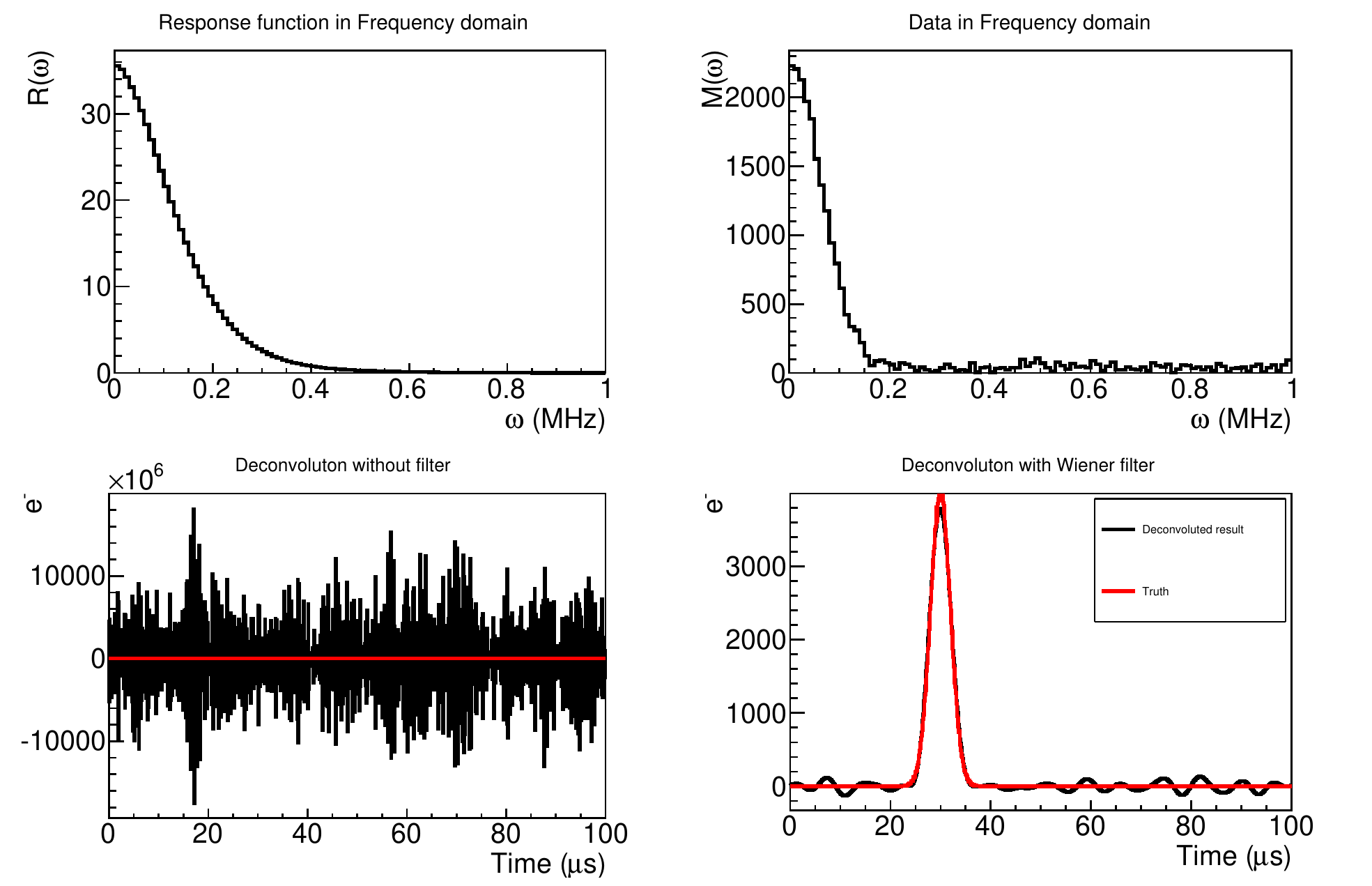}
\caption{(Top left) Response function $R(\omega)$ and (top right) data $M(\omega)$
  are shown for the example in Fig.~\ref{fig:simu_signal}.
  Deconvoluted results with (bottom right panel) and without (bottom left panel) 
  Wiener filter are compared with truth.}
\label{fig:decon}
\end{figure}

With a suitable noise filtering model, an improved estimator for the signal
$\hat{S}(t)$ in the time domain can then be found by applying an inverse Fourier 
transform to $\hat{S}(\omega)$.  Essentially, the deconvolution replaces the real field and 
electronics response function ($R$) with an effective filter response function 
($F$ as in the right panel of Fig.~\ref{fig:wiener}). For the example shown in
Fig.~\ref{fig:simu_signal}, the response function in frequency domain $R(\omega)$
and the measured data in frequency domain $M(\omega)$ are shown in top left and
top right panel of Fig.~\ref{fig:decon}, respectively. The deconvoluted results
without (left panel with Eq.~\eqref{eq:decon_2}) and with (right panel with
Eq.~\eqref{eq:decon_filt}) the Wiener filter are shown in the bottom left and the
bottom right panel of Fig.~\ref{fig:decon}. Without the (Wiener) filter,
the noise in the measured data at high frequency is significantly amplified by
dividing the small value of response function, which leads to unacceptable 
fluctuations in the deconvoluted results. With the Wiener filter applied,
the deconvoluted results are comparable to the simulated signal truth. 
Since the deconvolution problem shares many common features with the data unfolding
problem, it is natural to extend the application of Wiener filter technique
from deconvolution to unfolding.

\section{SVD Unfolding with Wiener Filter}\label{sec:algorithm}

\begin{table}[h]
\begin{center}
\begin{tabular}{lll}
\hline
\hline
Symbols & Meaning & Dimension and Format  \\
\hline
$A$ & Constructed smearing matrix from $W$ and $V$  & $n\times{}n$ matrix \\
$A_{C}$ & Constructed smearing matrix with $C$ & $n\times{}n$ matrix \\
$C$ & Assisting matrix, i.e. 1$^{st}$ or 2$^{nd}$ derivative matrix &  $n\times{}n$ matrix \\
$Cov$ & Covariance matrix of measurement ${\bf m}$ & $m\times{}m$ matrix \\
$Cov_{\hat{s}}$ & Covariance matrix of unfolded result $\hat{s}$& $n\times{}n$ matrix \\
$D$   & Center diagonal matrix from SVD decomposition of $R$ & $m\times{}n$ matrix \\
$D_C$   & Center diagonal matrix from SVD decomposition of $R \cdot C^{-1}$ & $m\times{}n$ matrix \\
$d_{i}$ & Non-negative diagonal element of $D$, $d_{i}$ = $D_{ii}$  &  \\
$F$ & regularization filter & $n\times n$ matrix \\
${\bf m}$  & Measured spectrum & $m$ vector \\
$M$  & Measured spectrum after pre-scaling & $m$ vector \\
$M_U$  & $U^{T} \cdot M$ & $m$ vector \\
$\overline{M_U}$  & Expectation of $U^{T} \cdot M$ based on $\overline{s}$& $m$ vector \\
$N$   & ``Noise'' of measurement after pre-scaling & $m$ vector \\
$N_{U}$   & $U^{T}\cdot N$ & $m$ vector \\
$Q$  & Lower triangular matrix from Cholesky decomposition of $Cov^{-1}$ & $m\times{}m$ matrix \\
${\bf r}$  & Smearing/response function  & $m\times{}n$ matrix \\
$R$  & Smearing/response matrix after pre-scaling & $m\times{}n$ matrix \\
$R_{tot}$ & Total transformation matrix connecting $\hat{s}$ and ${\bf m}$ & $n\times{}n$ matrix \\
$s$  & Unknown spectrum (a variable in the $\chi^2$ calculation)   & $n$ vector \\
$s_{true}$  & True spectrum & $n$ vector \\
$\overline{s}$  & Expectation of true spectrum & $n$ vector \\
$\hat{s}$  & Unfolded spectrum & $n$ vector \\
$T_{bias}$ & Bias of unfolded result &  $n$ vector \\
$T_{deviation}$  & Deviation (square-root of variance) of unfolded result & $n$ vector \\
$T_{deviation~j}$  & $j$th element of deviation  &  \\
$U$   & Left U matrix from SVD decomposition of $R$ & $m\times{}m$ matrix \\
$U_{C}$ & Left matrix from SVD decomposition of $R\cdot{}C^{-1}$ & $m\times{}m$ matrix \\
$V^{T}$  & Right $V^{T}$ matrix from SVD decomposition of $R$ & $n\times{}n$ matrix \\
$V^{T}_{C}$  & Right matrix from SVD decomposition of  $R\cdot{}C^{-1}$ & $n\times{}n$ matrix \\
$W$ & Wiener filter & $n\times{}n$ matrix \\
$W_{C}$ & Wiener filter with  $C$ & $n\times{}n$ matrix \\
\hline
\end{tabular}
\caption{Explanations of main symbols used in this section.}
\label{tab:eq_def}
\end{center}
\end{table}

In this section, we describe the procedure of Wiener-SVD unfolding.
For clarity, Tab.~\ref{tab:eq_def} summarizes the symbols used in this section.

\subsection{Problem definition with SVD decomposition}

The data unfolding problem generally starts with a $\chi^2\left( s\right)$ function defined as
\begin{eqnarray}\label{eq:chi2}
  \chi^2\left( s\right) = \left( {\bf m} - {\bf r} \cdot s \right)^T \cdot Cov^{-1} \cdot \left( {\bf m} - {\bf r}\cdot s\right).
\end{eqnarray}
Here, ${\bf r}$ is an $m$ (row) $\times$ $n$ (column) smearing matrix that
connects the vector of measured data ${\bf m}$ (an m-dimensional vector)
with an unknown vector of signal $s$ (an n-dimensional vector). Note, we later use
$s_{true}$ to represent the true signal in order to differentiate from $s$ which is
a variable in calculating $\chi^2$. This matrix
is general and not limited by functional format in Eq.~\ref{eq:sym_condition}.
We use $\hat{s}$ to represent the estimator of the true signal $s_{true}$, which
is obtained after minimizing the chi-square function. We further restrict
ourselves in $m\ge n$ case. The matrix $Cov$ is an $m \times m$ covariance
matrix containing all statistical and systematic uncertainties associated
with ${\bf m}$ and ${\bf r}$ in calculating the differences between actual
measurement ${\bf m}$ and the expectation ${\bf r}\cdot s$. For example,
the covariance matrix would include i) statistical uncertainties from data,
ii) statistical and systematic uncertainties for the backgrounds, and iii)
statistical (with Monte Carlo simulation) and systematic uncertainties
associated with the detector response ${\bf r}$.~\footnote{In order to evaluate
  the systematic uncertainties associated with the detector response, one typically
  runs many Monte Carlo simulations with different detector responses to
  calcualte the expectated spectra. These spectra are compared with the nominal-detector-response
spectrum to construct covariance matrix.}

Since the covariance matrix $Cov$ is symmetric, the inverse
of it ($Cov^{-1}$) is also symmetric. Hence, $Cov^{-1}$ can be decomposed
with Cholesky decomposition into 
\begin{equation}
  Cov^{-1} = Q^T \cdot Q,
\end{equation}
where $Q$ is a lower triangular matrix and $Q^{T}$ is its transpose.  

Eq.~\ref{eq:chi2} can then be rewritten as 
\begin{equation}\label{eq:chi2_1}
  \chi^2 = \left(M -R \cdot s \right)^T \cdot \left(M -R \cdot s \right)
  = \sum_i \left(M_i -\sum_j R_{ij} \cdot s_j \right)^2,
\end{equation}
with $M := Q\cdot {\bf m}$ and $R := Q \cdot {\bf r}$, and this process is
commonly referred to as pre-scaling or pre-whitening. 
The solution after minimizing Eq.~\ref{eq:chi2_1} would be
$M = R \cdot \hat{s}$ or 
\begin{equation}\label{eq:sol_1}
  \hat{s} = \left( R^T R \right)^{-1} \cdot R^T \cdot M.
\end{equation}

In analogy to Eq.~\ref{eq:decon_2p}, Eq.~\ref{eq:sol_1} can be rewritten as
\begin{equation}\label{eq:sol_2}
\hat{s} = \left( R^T R \right)^{-1} R^T \cdot \left( R\cdot s_{true} + N \right). 
\end{equation}
with $N$ representing the ``noise'' coming from uncertainties (statistical
and systematic uncertainties associated with both ${\bf m}$ and ${\bf r}$).
Since $N=M- R\cdot s_{true} = Q \cdot \left( {\bf m} - {\bf r}\cdot s_{true}\right)$,
each term in the noise vector after pre-scaling follows a normal distribution
with $\mu=0$ and $\sigma=1$, since the denominator of the chisquare function
in Eq.~\ref{eq:chi2_1} (i.e. square of error) is unity. Given the fact that
each term in the noise vector is independent (i.e. uncorrelated), we refer
the basis in this domain to be orthogonal.

Using the singular value decomposition (SVD) approach, $R$ can be decomposed as
\begin{equation}\label{eq:svd}
  R = U \cdot D \cdot V^T,
\end{equation}
with both $U$ ($m \times m $) and $V$ ($n \times n$) being orthogonal matrices that satisfy
$U^T \cdot U = U \cdot U^T = I_{(m\times m)}$ and $V^T \cdot V = V \cdot V^{T} = I_{(n\times n)}$.
$I$ is the identity matrix and the subscript represents the dimension.  
$D$ is an $m \times n$ diagonal matrix with non-negative diagonal elements
(known as singular values) $D_{ii}$ = $d_{i}$ arranged in descending order as
$i$ increases.   

Inserting Eq.~\ref{eq:svd} into Eq.~\ref{eq:sol_2}, we have
\begin{eqnarray}\label{eq:sol_3}
  \hat{s} &=& V \cdot D^{-1} \cdot U^{T} \cdot \left( R\cdot s_{true} + N \right) \nonumber \\
  &=& V \cdot D^{-1} \cdot \left( R_U \cdot s_{true} + N_U \right) \nonumber \\ 
  &=& V \cdot D^{-1} \cdot M_U.
\end{eqnarray}
where $R_U := U^T \cdot R$, $N_U := U^T \cdot N$, and $M_U := U^{T} \cdot M$
are transformations of the smearing matrix, the noise $N$, and the measured signal, respectively.
This equation can be compared to Eq.~\ref{eq:decon_2}. 
Note, since $U$ is an orthogonal matrix and elements of the
original noise vector $N$ are uncorrelated, elements of the new noise
vector $N_U$ are still uncorrelated. Each element follows a normal
distribution with $\mu=0$ and $\sigma=1$. Thus, the basis in this new
domain is still orthogonal.

Given Eq.~\ref{eq:sol_3}, we can understand the large fluctuation in the
unfolded results. First, due to the existence of the smearing in matrix ${\bf r}$ or
$R$, the magnitude of singular values $d_{i}$ drops significantly as $i$
increases.  After a certain $i$, the value of $d_{i}$ can be extremely small which
leads to a gigantic value in the corresponding element in $D^{-1}$.
In the case of perfect signal without any noise $N_U$, these gigantic diagonal
elements are effectively canceled out by the small values in the signal
$R \cdot s$ leading to recover the signal without any bias after data unfolding.
The situation is completely changed in the presence of noise $N_U$.
Since these noise will be significantly amplified after multiplying with $D^{-1}$,
the unfolded results suffer from large fluctuations. From the above discussions,
it is easy to see the similarities between Eq.~\ref{eq:sol_3} and Eq.~\ref{eq:decon_2}
for the deconvolution discussion in Sec.~\ref{sec:introduction}.
Therefore, in analogy to the results after the Fast Fourier
Transformation (FFT), we refer to $M_U$ after the SVD transformation
as the measurement in the {\it effective frequency domain} in analogy to
the frequency domain in the signal processing (e.g. Eq.~\ref{eq:decon_2p}
and Eq.~\ref{eq:decon_2}). Both these domains have orthogonal basis that
are linear transformation from the original basis.
While FFT requires the functional format of response function to be
symmetric (i.e. Eq.~\ref{eq:sym_condition}), the SVD does not require
this symmetric condition and is more general.

\subsection{Review of traditional regularization approach}\label{sec:review_regularization}

Regularization is a commonly used technique to address the problem described in
the previous section. It imposes additional constraints on the estimation of
true distribution $s$ by introducing a regularization function $\Sigma(s)$. The estimator can be
obtained by finding the maximum of a weighted combination of log-likelihood $\log{L}$ and
$\Sigma$:
\begin{equation}\label{eq:regularization}
\phi(s)=\log{L(s)}+\tau \Sigma(s),
\end{equation}
where $\tau$ is called regularization strength, which determines the
trade-off between bias due to imposed constraints and variance due to existence
of the noise in the unfolded distribution. In general, to obtain the best estimation of the
signal, the log likelihood $\log{L(s)}$ as well as the regularization function $\Sigma(s)$
are required to be sufficiently well-behaved (e.g. at the very least
  there should not contain multiple local maxima~\cite{cowan_book}). 

The so-called \textbf{Tikhonov regularization} technique uses the following
regularization function
\begin{equation}\label{eq:tikhonovfunction}
  \Sigma^{k}\left(s(E)\right)=-\int \left( \frac{\text{d}^{k}s(E)}{\text{d}^{k}E} \right)^2~\text{d}E, k = 0, 1,2,3,\cdots,
\end{equation}
with the spectrum $s$ depending on the variable $E$ (e.g. energy).
For example, when $k=0$, we have $\Sigma^{0}\left(s(E)\right)=-\int \left( s(E) \right)^2
\text{d}E$, which favors small values of the signal. Another commonly used example is
$k=2$, in which $\Sigma^{2}\left(s(E)\right)$ represents a measure of
the average curvature of distribution $s(E)$, imposing a constraint on the
smoothness of the signal. When applying the regularization technique, one generally
chooses a regularization function, evaluates the bias and variance of the estimator
as a function of regularization strength $\tau$. The value of $\tau$ is then optimized
based on a predetermined metric.

  In the case of $\Sigma^{0}\left(s(E)\right)$ regularization, the unfolded results can be
  expressed as:
\begin{eqnarray}\label{eq:reg1}
  \hat{s} &=& A \cdot \left( R^T R \right)^{-1} \cdot R^T \cdot M  \nonumber \\
  & = & A \cdot V \cdot D^{-1} \cdot \left( R_U \cdot s_{true} + N_U \right).
\end{eqnarray}
Here $A$ behaves as an additional smearing matrix added to the unfolding results of Eq.\ref{eq:sol_3}. It has the form of:  
\begin{eqnarray}\label{eq:additional}
  A = V \cdot F \cdot V^T, 
\end{eqnarray}
where $F$ is an $n\times n$ diagonal matrix with elements satisfying
\begin{eqnarray}\label{eq:wiener_form}
F_{ii} = \frac{d_{i}^{2}}{d_{i}^{2} + \tau }  
\end{eqnarray}
here $\tau$ is the regularization strength. Eq.~\ref{eq:reg1} is then changed to
\begin{eqnarray}\label{eq:reg2}
  \hat{s} & = & V \cdot F \cdot D^{-1} \cdot \left( R_U \cdot s_{true} + N_U \right).
\end{eqnarray}
At small values of $d_i$, the corresponding noise terms in $N_U$ are now suppressed
by $d_i/\left(d_i^2 + \tau \right)$ at finite value of $\tau$ instead of being
amplified by $1/d_i$. Such a change would suppress the large fluctuations
in the unfolded results due to the presence of the noise $N_U$. It's worth noting that
the regularization method is effectively introducing an additional smearing to the unfolding results,
which would lead to biases on the unfolded signal. The above derivation is similar for other
$\Sigma^{k}\left(s(E)\right)$ regularization schemes. 

\subsection{Wiener-SVD approach}\label{sec:wiener_svd}

In the $\Sigma^{0}\left(s(E)\right)$ regularization, the functional form of
$F_{ii}= d_i^2 / \left( d_i^2 + \tau \right)$ is independent of
the signal shape, and the choice of $\tau$ is obtained through a scan of this parameter
with respect to some metrics. With the concept of the Wiener filter, we can construct
$W$ (replacing $F$ in regularization) directly to optimize the signal to noise ratio.
This replacement defines the Wiener-SVD approach, in which the functional form of
$W$~\footnote{We replace $F$ by $W$ for Wiener filter.}
also considers the expectation value of the signal in the effective frequency domain:~\footnote{In general, the $s_{true}$ is unknown, so the expectation of signal $\overline{s}$ is used.}
\begin{eqnarray}\label{eq:signal_expectation}
  \overline{M_U} = U^{T} \cdot \overline{M} 
   =  U^{T} \cdot R \cdot \overline{s} 
   =   D \cdot V^T \cdot \overline{s}.
\end{eqnarray}
Here, the expectation of signal is assumed to be known. We will come back to this point
later in this section on how to obtain the expectation of signal.
The construction of $W$ is based on the Wiener filter 
${\rm``}\overline{R^2\cdot S^2}{\rm"}/\left({\rm``}\overline{R^2 \cdot S^2}{\rm"} + {\rm``}\overline{N^2}{\rm"}\right)$ as
in Eq.~\ref{eq:wiener}.
Taking Eq.~\ref{eq:signal_expectation}, at bin $i$, we have
\begin{eqnarray}
  {\rm``}\overline{R^2 \cdot S^2}{\rm"} &=& \overline{M}^2_U= d_i^2 \cdot \left( \sum_j V^T_{ij} \cdot \overline{s}_j \right)^2 \label{eq:signal_exp} \\
  {\rm``}\overline{N^2}{\rm"} &=& 1, \label{eq:noise}
\end{eqnarray}
resulting in a Wiener filter of
\begin{equation}\label{eq:wiener_reg}
  W_{ik}  = \frac{d_i^2 \cdot \left( \sum_{j} V^T_{ij} \cdot \overline{s}_j \right)^2}{d_i^2 \cdot \left( \sum_{j} V^T_{ij} \cdot \overline{s}_j \right)^2+1} \cdot \delta_{ik},
\end{equation}
replacing $F$ in Eq.~\ref{eq:additional}.
Here, Eq.~\ref{eq:noise} is obtained, since each element of noise $N_U$
follows a normal distribution with $\mu=0$ and $\sigma=1$.
We have 
\begin{equation}
  \left\{W \cdot D^{-1}\right\}_{ij} = \frac{d_i \cdot \left( \sum_k V^T_{ik} \cdot \overline{s}_k \right)^2}{ \left( d_i^2 \cdot \left( \sum_k V^T_{ik} \cdot \overline{s}_k \right)^2 + 1 \right)} \cdot \delta_{ij}.
\end{equation}
The small value of $d_i$ is balanced by the finite value of the expectation value
of $\overline{N^2}\equiv1$. From Eq.~\ref{eq:wiener_reg}, the construction of the Wiener
filter takes into account both the strengths of signal and noise expectations and is
free from regularization strength $\tau$.

A few comments should be made regarding the Wiener-SVD approach:
\begin{itemize}
\item {\bf Generalized Wiener-SVD approach:} \\
  As shown in Ref.~\cite{Hocker:1995kb}, the regularization can be applied on the
  curvature of the spectrum instead of the strength of the spectrum. This involves
  an additional matrix $C_2$. This can also be achieved in the Wiener-SVD approach:
\begin{equation}
  \overline{M} = R \cdot C^{-1} \cdot C \cdot \overline{s}
\end{equation}
by including an additional matrix $C$ that has the commonly used regularization forms,
such as the first and second derivatives. Since the effective frequency domain is
determined by the smearing matrix $R$, the inclusion of $C$ would alter the basis
of the effective frequency domain.
In this case, the SVD decomposition becomes
\begin{equation}
R \cdot C^{-1} =  U_C \cdot D_C \cdot V^T_{C}. 
\end{equation}
The final solution of the regularization would become
\begin{equation}\label{eq:wiener_unfold}
  \hat{s} = C^{-1} \cdot V_C \cdot W_C \cdot V^T_C \cdot C \cdot
  \left( R^T R \right)^{-1} \cdot R^T \cdot  M.
\end{equation}
or
\begin{equation}\label{eq:wiener_unfold2}
\hat{s} = A_{C} \cdot (R^{T}R)^{-1}\cdot R^{T} \cdot M,
\end{equation}
where
\begin{equation}\label{eq:wiener_unfold3}
 A_{C} =  C^{-1} \cdot V_C \cdot W_C \cdot V^T_C \cdot C.
\end{equation}
The corresponding Wiener filter would be
\begin{equation}\label{eq:wiener_reg1}
  W_{ii}  = \frac{d_{Ci}^2 \cdot \left( \sum_j V^T_{Cij} \cdot \left(\sum_l C_{jl} \cdot \overline{s}_l \right)\right)^2}{d_{Ci}^2 \cdot \left( \sum_j V^T_{Cij} \cdot \left( \sum_l C_{jl}\cdot \overline{s}_l \right)\right)^2+1},
\end{equation}
where $C_{jl}$, $V^T_{Cij}$, and $d_{Ci}$ are matrix elements
of matrices $C$, $V_C$, and $D_C$, respectively.
\item {\bf Covariance matrix of unfolded results:} \\
  Since the unfolded results are a linear transformation of the measurement, we
  can easily evaluate the uncertainties associated with them. 
  Eq.~\ref{eq:wiener_unfold2} can be rewritten into
  \begin{equation}\label{eq:sunfold}
    \hat{s} = R_{tot}\cdot {\bf m}
  \end{equation}
  with
  \begin{equation} \label{eq:rtot}
    R_{tot} = A_{C} \cdot
    \left( R^T R \right)^{-1} \cdot R^T \cdot Q.
  \end{equation}
  Then, the covariance matrix of $\hat{s}$ can be deduced from the $Cov$ (the covariance
  matrix of $M$) as
  \begin{equation}\label{eq:unfold_cov}
    Cov_{\hat{s}} = R_{tot} \cdot Cov \cdot R^{T}_{tot}.
  \end{equation}
  
\item {\bf Variance:} \\
  The variances of the unfolded data can also be easily calculated given
  that their origin $N$ in Eq.~\ref{eq:wiener_unfold} is well understood.
  Defining $N\left(i\right)$ as a vector with the $i$th element
  being 1 and the rest of elements being 0, we can calculate the variance
  in $s$ due to $i$th element in $N$ as:
  \begin{equation}\label{eq:var}
    T_{deviation}\left(i\right) = A_{C} \cdot
    \left( R^T R \right)^{-1} \cdot R^T \cdot N\left(i\right),
  \end{equation}
  with $T_{deviation}\left( i\right)$ being a vector. 
  The variance of the $j$th element of $T_{deviation~j}$ can thus be written as:
  \begin{equation}
    T_{deviation~j} = \sqrt{\sum_i T^2_{deviation~j}\left(i\right)},
  \end{equation}
  after summing the contribution from each independent noise source.
  The square of $T_{deviation~j}$ corresponds to the $j$th diagonal element
  of covariance matrix in Eq.~\ref{eq:unfold_cov}.
\item {\bf Bias:} \\
  Given Eq.~\ref{eq:wiener_unfold2}, we can understand the entire process of
  unfolding as to "remove" the effect of $R$ through multiplying
  $\left( R^T R \right)^{-1} \cdot R^T$ and then replace it with a new
  smearing matrix $A_{C}$. Therefore, it is straightforward to estimate the
  bias on the unfolded results:
  \begin{eqnarray}\label{eq:bias}
    T_{bias} &=& \left( A_{C} -I\right) \cdot
    \left( R^T R \right)^{-1} \cdot R^T \cdot \overline{M},\\
      &=& \left(A_{C} - I\right)\cdot \overline{s}, \nonumber
  \end{eqnarray}
  with $I$ being identity matrix and $\overline{s}$ being the
  expectation of the signal.

  Given only the measurement {\bf m}, an alternative approach given in Ref.~\cite{cowan_review} can be used to estimate the bias. The bias for bin $i$ now is defined as:
\begin{eqnarray}\label{eq:bias2}
  T_{bias}'\left(i\right) = \sum_{j}\frac{\partial{}\hat{s}\left(i\right)}{\partial{}M\left(j\right)}\cdot{}(\hat{M}\left(j\right) - M\left(j\right)),
\end{eqnarray}
where $\hat{M} = R \cdot \hat{s}$. Using Eq.~\ref{eq:wiener_unfold2} and ~\ref{eq:bias2}, we have 
\begin{eqnarray}\label{eq:bias3}
  T_{bias}' = A_C \cdot \left( A_C - I \right)\cdot\left( R^T R \right)^{-1} \cdot R^T \cdot M  = \left(A_{C} - I \right)\cdot \hat{s}. 
\end{eqnarray}
As can be seen, in Eq.~\ref{eq:bias} the bias is directly calculated if $\overline{s}$ is known, while for Eq.~\ref{eq:bias3} the bias is estimated by replacing the $\overline{s}$ in Eq.~\ref{eq:bias} with $\hat{s}$. For Tikhonov regularization,
the bias estimation of Eq.~\ref{eq:bias3} deviates from that of
Eq.~\ref{eq:bias} at large $\tau$ values when the unfolded spectrum
significantly deviates from the true spectrum. At very small values of $\tau$,
the unfolded spectrum suffers from large fluctuations leading to a significant
overestimation of bias with Eq.~\ref{eq:bias3}.
Moreover, since $M$ has fluctuations,
the variance of $T_{bias}'$ can also be calculated.
The variance in $T_{bias}'$ due to $i$th element in $N_U$ has the form of:
\begin{equation}\label{eq:bias_var}
  T_{deviation}^{b}\left(i\right) =  \left(A_{C} - I \right) \cdot A_{C} \cdot
  \left( R^T R \right)^{-1} \cdot R^T \cdot N\left(i\right).
\end{equation}

\item {\bf Expectation of signal:} \\
  We should note that Eq.~\ref{eq:signal_exp} only considers one
  particular model of signal $s$. In reality, the expectation of signal should
  cover a range of possible signals $\overline{s(k)}$ (e.g. prior models) that
  are compatible with existing observations
\begin{equation}\label{eq:signal_exp1}
{\rm``}\overline{R^2\cdot S^2}{\rm"} = \overline{M}^2_U = d_{Ci}^2 \cdot \frac{ \sum_k \left( \sum_j V^T_{Cij} \cdot \left( \sum_{l} C_{jl} \cdot \overline{s(k)}_l \right) \right)^2 \cdot e^{-\frac{\chi_k^2}{2}}}{\sum_k e^{-\frac{\chi_k^2}{2}}}.
\end{equation}
  
  Here, $\chi_k^2$ is the chi-square representing the compatibility
  between the prediction $s(k)$ and the measurement. When there is no prior models
  available, one can construct these models using general functions (e.g.
  Legendre polynomials or spline functions). 
\item {\bf Regularization interpretation of Wiener-SVD approach:} \\ 
  We show that the Wiener-SVD
  unfolding method is equivalent to a regularization which attempts to maximize the
  signal to noise ratio in
  the effective frequency domain $M_{U~i}=D_i \cdot \sum_{j} V^{\intercal}_{ij}s_{j}$.
  Recall Eq.~\ref{eq:regularization}, one now has:
  \begin{equation}
    \phi(s)=\log{L(s)}+\frac{1}{2}\sum_{i}\log{\frac{M^2_{U~i}}{\overline{N^2}}} = \log{L(s)}+\frac{1}{2}\sum_{i}\log{\frac{\left(D_i \cdot \sum_{j} V^{\intercal}_{ij} \cdot s_{j}\right)^2}{1}}
  \end{equation}
  with the expectation of noise square being $1$ in the effective frequency domain.
  Using the procedure detailed by~\cite{cowan_book}, by maximizing $\phi(s)$ one obtains the
  following estimator
  \begin{equation}
    \hat{s}=-X^{-1}\cdot Y \cdot {M_U}
    \label{eq:ABestimator}
  \end{equation}
  where
  \begin{align}
    X_{ij}&=\frac{\partial^2\phi^2}{\partial s_{i}\partial s_{j}}=-({R}^\intercal {R})_{ij}-\sum_{k}V_{ik}\cdot \frac{1}{M^2_{U~k}} \cdot V_{kj}^\intercal \\
    Y_{ij}&=\frac{\partial^2\phi^2}{\partial s_{i}\partial M_{U~j}}={R}^\intercal_{ij}
  \end{align}
  With $X$ and $Y$ evaluated at the expectation of $\overline{s}$ and $\overline{M}$,
  Eq.\ref{eq:ABestimator} can be rewritten as
  \begin{equation}
    \hat{s}=V \cdot (D^2+{D'}^2)^{-1} \cdot V^\intercal \cdot {R}^\intercal \cdot \overline{M}
  \end{equation}
  where 
  \begin{equation}
    {D'}_{ij}=\frac{1}{\overline{M_{U~i}^2}} \cdot \delta_{ij}.
  \end{equation}
  Therefore, we have
  \begin{equation}
    A = V \cdot (D^2+{D'}^2)^{-1} \cdot D^2\cdot V^\intercal = V \cdot W \cdot V^\intercal,
  \end{equation}
  with
  \begin{equation}
    W_{ij}=\frac{d_i^2}{d_i^2+\frac{1}{\overline{M^2_{U~i}}}}\delta_{ij} = \frac{d_i^2 \cdot \left( \sum_j V^T_{ij} \cdot \overline{s}_j \right)^2}{d_i^2 \cdot \left( \sum_j V^T_{ij} \cdot \overline{s}_j \right)^2+1} \cdot \delta_{ij}
  \end{equation}
  One recovers Eq.\ref{eq:wiener_reg}.
\end{itemize}

\section{Data Unfolding Example: Cross Section Extraction}\label{sec:xs}
%
%
%

In this example, we apply the Wiener-SVD unfolding on a neutrino cross section
extraction problem. As introduced in Sec.~\ref{sec:introduction}, the data unfolding
technique can be useful for this problem when the ratio of cross sections is desired
or when the comparison of cross section measurements from different experiments is needed.

\begin{figure}
\includegraphics[width=1.0\textwidth]{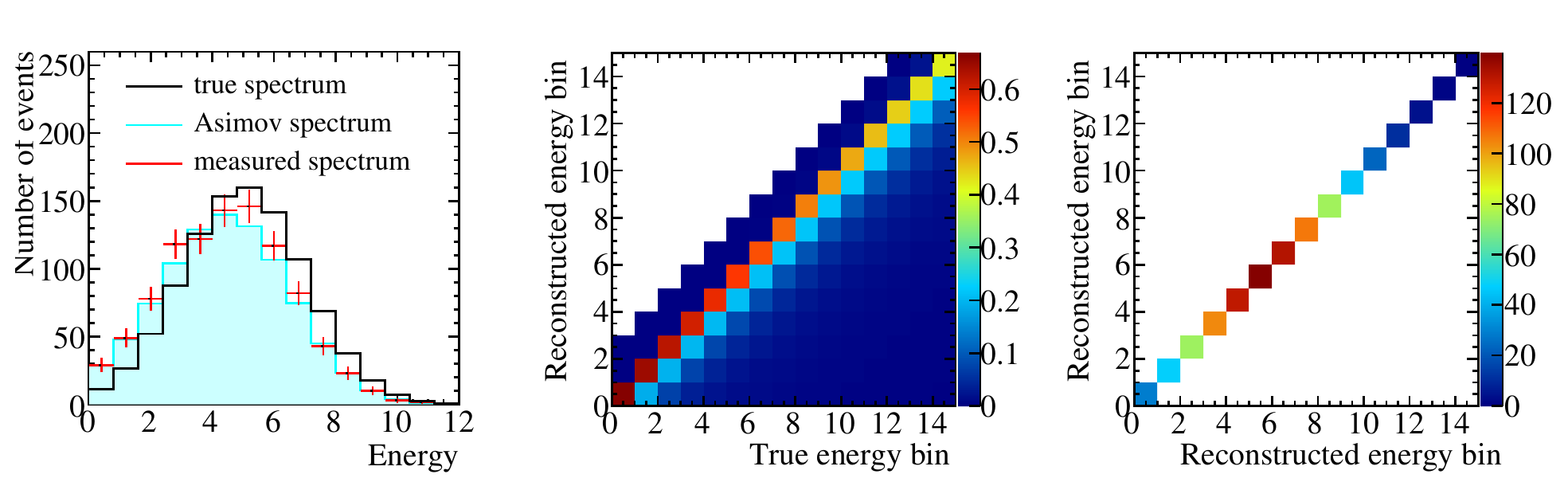}
\caption{(Left) True energy spectrum $s_{true}$ (black), Asimov data
  spectrum \cite{cowan:2011likelihoodtest} ${\bf r}\cdot s_{true}$(cyan),
  and measured spectrum ${\bf m}$ (red)
  are shown. (Middle) Detector smearing matrix ${\bf r}$ is shown. (Right) Covariance matrix $Cov$
  (statistical only) is shown.}
\label{fg:xs1}
\end{figure}

Experiments that engage in neutrino cross-section measurements generally consist of
two parts: a neutrino beam produced by bombarding a target with a proton
beam, and a detector (or a series of detectors) located a few hundred meters away from the target to detect neutrino interactions.
The neutrino beam composition and energy distribution are generally well-understood.
Depending on the detector technology, the neutrino energy can be reconstructed via
calorimetry or from the kinematics of final state particles. Due to the smallness of neutrino cross-sections,
the signal statistics are typically low in such measurements. Depending on the beam configuration 
(on-axis or off-axis) one can have a broad or narrow neutrino energy spectrum. Some neutral hadrons
produced by neutrino interactions, neutrons in particular, could leave undetectable for calorimetric or tracking
detectors, therefore the reconstructed visible energy tends to be smaller than the true neutrino energy.
For simplicity, we neglect the neutrino flux uncertainties and only consider the reconstructed neutrino energy spectrum $M(E_{r})$. 
Figure~\ref{fg:xs1} shows the true energy spectrum $s_{true}$ , detector smearing
matrix ${\bf r}$, Asimov spectrum~\cite{cowan:2011likelihoodtest} $\overline{\textbf m}={\textbf
  r} \cdot s_{true}$, measured spectrum $\textbf{m}$ which is the
Asimov spectrum with random Poisson statistical fluctuation, and
covariance matrix with statistical uncertainty only. 
A Gaussian true spectrum is assumed; detector smearing matrix is mocked up such that reconstructed energy is skewed towards energy lower than the true neutrino energy.
No systematic uncertainty or background is considered in this toy experiment.


In order to illustrate the performance of the Wiener-SVD method, we compare the unfolded results
with those from the Tikhonov regularization described in Sec.~\ref{sec:review_regularization}.
Three choices of $C$ matrices are used for comparison. They are:
\begin{eqnarray}
  C_0 &=& \begin{bmatrix}
    1 & 0 & 0 & \dots & 0 \\
    0 & 1 & 0 & \dots & 0 \\
    \vdots & \vdots & \vdots & \ddots & \vdots \\   
       0  & 0 & 0 & \dots  & 1
  \end{bmatrix},
  C_1 = \begin{bmatrix}
    -1 & 1 & 0 & \dots &0 &0 \\
    0 & -1 & 1 & \dots &0 &0 \\
    \vdots & \vdots & \vdots & \ddots & \vdots & \vdots \\
     0  & 0 & 0 & \dots  & -1  & 1\\  
    0  & 0 & 0 & \dots  & 0 & -1
  \end{bmatrix}, \nonumber \\
  C_2 &=& \begin{bmatrix}
    -1+\epsilon & 1 & 0 & \dots & 0 & 0 & 0\\
    1 & -2+\epsilon & 1 & \dots & 0 & 0 & 0\\
    \vdots & \vdots & \vdots& \ddots & \vdots & \vdots & \vdots\\    
    0 & 0 &0  & \dots&1 &-2+\epsilon & 1 \\
    0  & 0 & 0 & \dots&0 & 1 & -1+\epsilon 
  \end{bmatrix},
\end{eqnarray}
which correspond to the k=0, k=1 (first-order derivative), and k=2 (second-order derivative, or curvature)
cases in Eq.~\ref{eq:tikhonovfunction}, respectively. Since $C^{-1}$ is needed to construct the Wiener
filter $W$, to make $C$ invertible, a very small value of $\epsilon=10^{-8}$ is added to the
diagonal elements of $C_{2}$ matrix. 

In addition, it should be noted
that one has the freedom to normalize the unfolded distribution to that of the measured
distribution. This is equivalent to imposing a constraint on the
total number of events in Eq.~\ref{eq:regularization}:
\begin{equation}\label{eq:regularization_ntot}
\phi(s)=\log{L(s)}+\tau\Sigma(s)+\lambda\left[\sum_{i=1}^N \overline{s}_i-\sum_{i=1}^{N}s_{i}\right],
\end{equation}
where $\lambda$ is a Lagrange multiplier, and $\partial\phi/\partial\lambda=0$. This normalization can be of particular importance for the $C_0$ case with low statistics or large systematic uncertainties. 

Given a choice of the matrix $C$, the unfolded results with the Wiener-SVD method can be obtained
through Eq.~\ref{eq:wiener_unfold}. For Tikhonov regularization, the unfolded
results also depend on the regularization strength $\tau$. For the results shown
in this section, the optimal regularization strength is determined by minimizing the following
Mean Square Error (MSE)~\cite{cowan_review}:
\begin{equation}\label{eq:reactor_mse1}
  {\rm MSE} = \frac{1}{n} \left(\sigma^2 +b^2 \right) = \frac{1}{n}\sum_{i}^{n}\left(T_{deviation~i}^{2}
  + T_{bias~i}^{2} \right),
\end{equation}
where $\sigma^{2} = \sum T_{deviation~i}^{2}$ is the total variance
and $b^{2} = \sum T_{bias~i}^{2}$ is the total bias square. Here $i$ represents the $i$th bin and
the definition of $T_{deviation}$ and $T_{bias}$ can be found in Eq.~\ref{eq:var} and Eq.~\ref{eq:bias},
respectively.

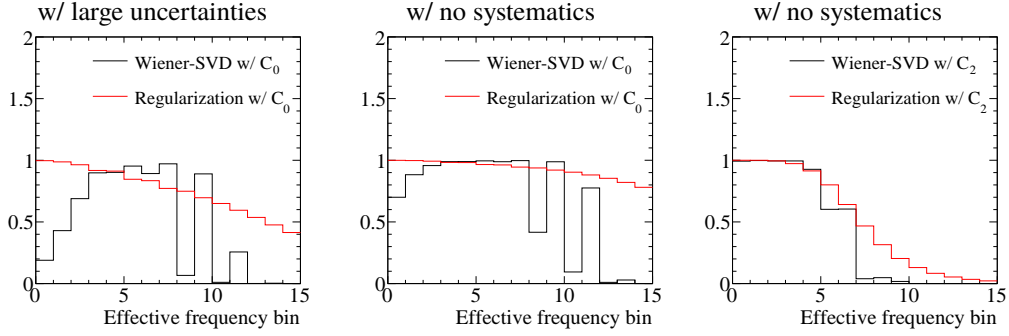
\begin{figure}
\begin{center}
\includegraphics[width=0.3\textwidth, trim=0cm 0cm 10cm 0cm, clip=true]{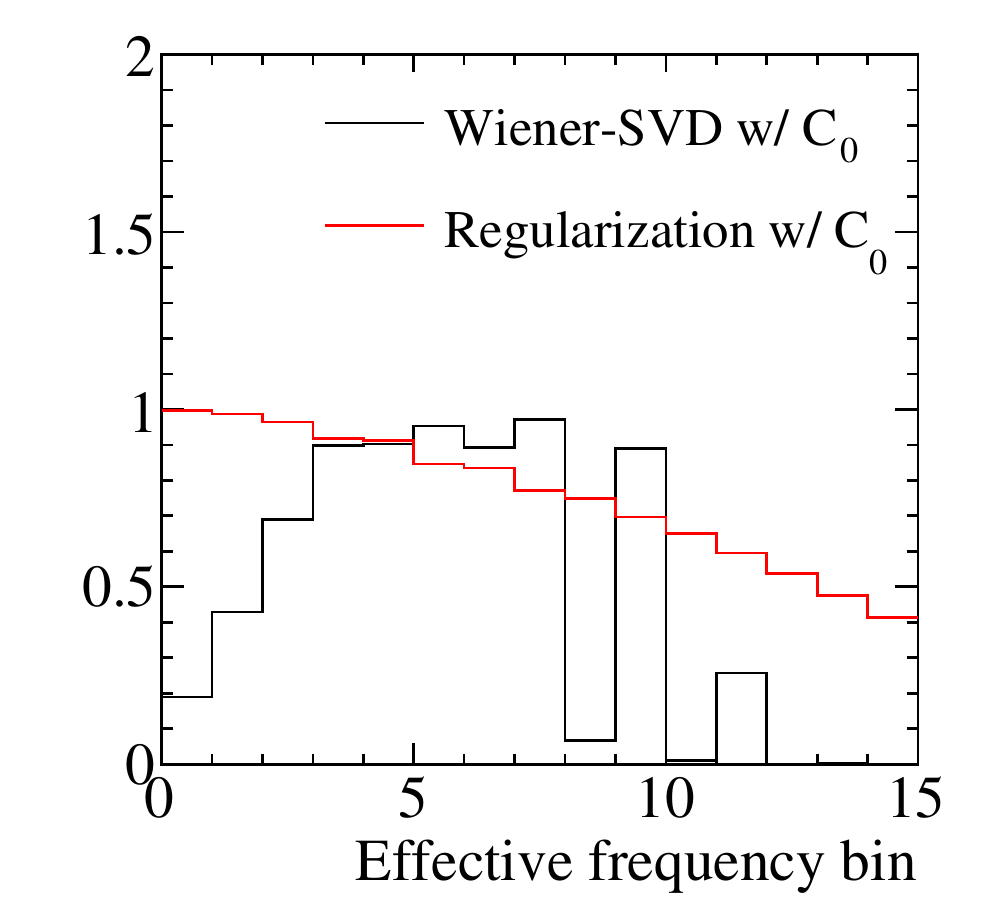}
\includegraphics[width=0.6\textwidth]{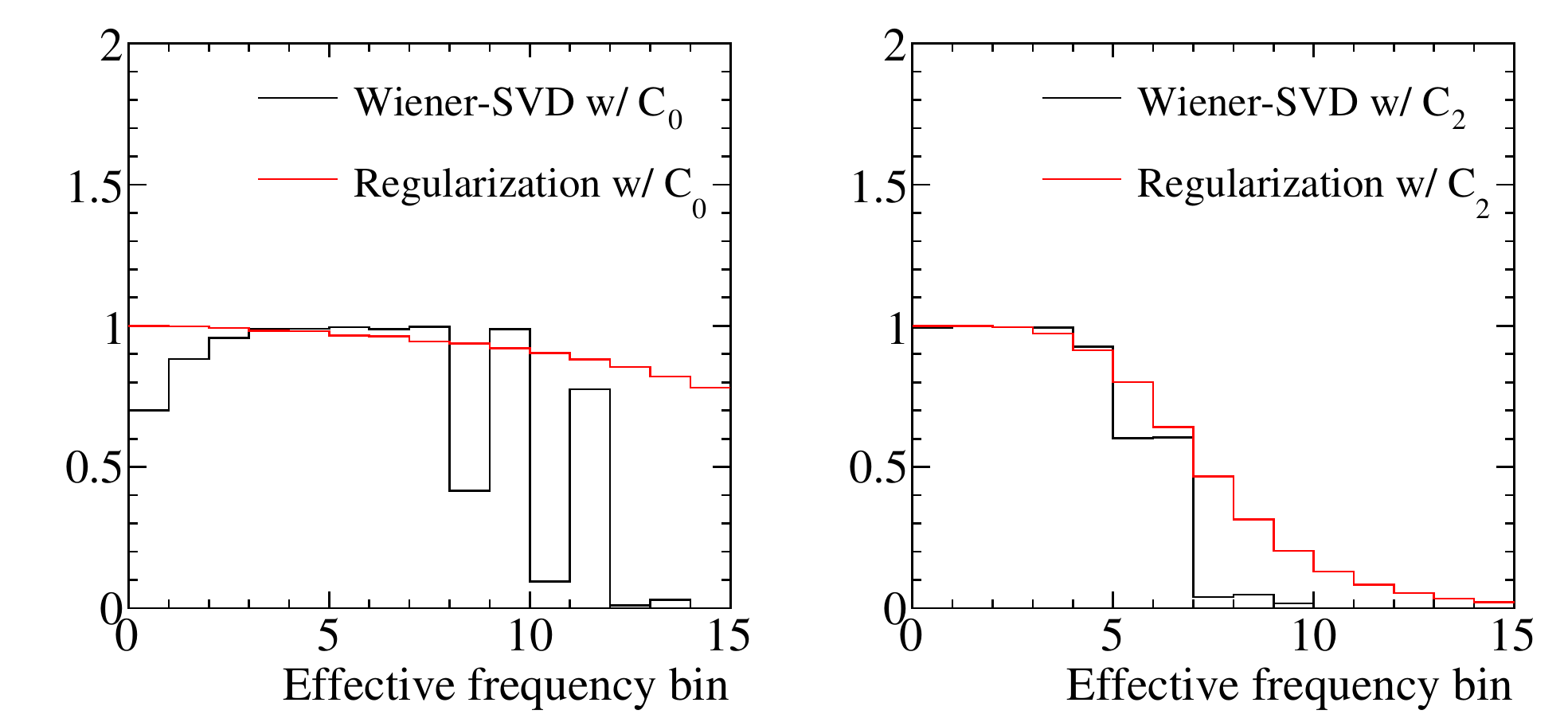}\\
\end{center}
\begin{picture}(0,0)
\put(42,145){\small w/ large uncertainties}
\put(180,145){\small w/ no systematics}
\put(310,145){\small w/ no systematics}
\end{picture}
\vspace{-5ex}\caption{Wiener filter $W$ and regularization filter $F$ in the effective frequency space.
  (Left) w/ $C_{0}$ and uncertainty (systematic and statistical) $\sqrt{10}$ times as big as shown Fig.~\ref{fg:xs1}.
  (Middle) w/ $C_{0}$ and default statistical-only uncertainty as shown in Fig.~\ref{fg:xs1}.
  (Right) w/ $C_{2}$ and default statistical-only uncertainty.
  The bins are ranked from low frequency (large eigenvalue of SVD) to high frequency
  (small eigenvalue of SVD). See text for more discussion.}
\label{fg:xs2}
\end{figure}

Figure~\ref{fg:xs2} shows the Wiener filter and regularization filter in the effective frequency
domain for the $C_0$ and $C_2$ cases. To construct the Wiener filter in this toy example, 
the expectation of signal $\overline{s}$ is taken to be the true signal $s_{true}$. We use
Eq.~\ref{eq:signal_exp1} to construct the Wiener filter in the example described in next section. 
The Wiener and regularization filters assign different weights
to each effective frequency bin: regularization is scaled by the eigenvalue corresponding to each bin,
whereas the Wiener filter is scaled by the \lq\lq{signal/noise-weighted}\rq\rq eigenvalue. Both the Wiener
and regularization filters suppress high frequency bins, and therefore reduce the impact from random
fluctuation at small $d_i$ values. In order to emphasize the importance of normalization, the left panel
of Fig.~\ref{fg:xs2} assumes that the total systematic uncertainty is $\sqrt{10}$ times as the statistical
uncertainty for the $C_0$ case. When the statistics are low, or the systematic uncertainty is high, the Wiener
 and regularization filters both yield greater suppression. Since the filter is multiplied
on the measurement, the normalization is often needed to further reduce the bias, especially when the
$C_0$ matrix is used.



\begin{figure} 
  \begin{center}
\includegraphics[width=0.35\textwidth]{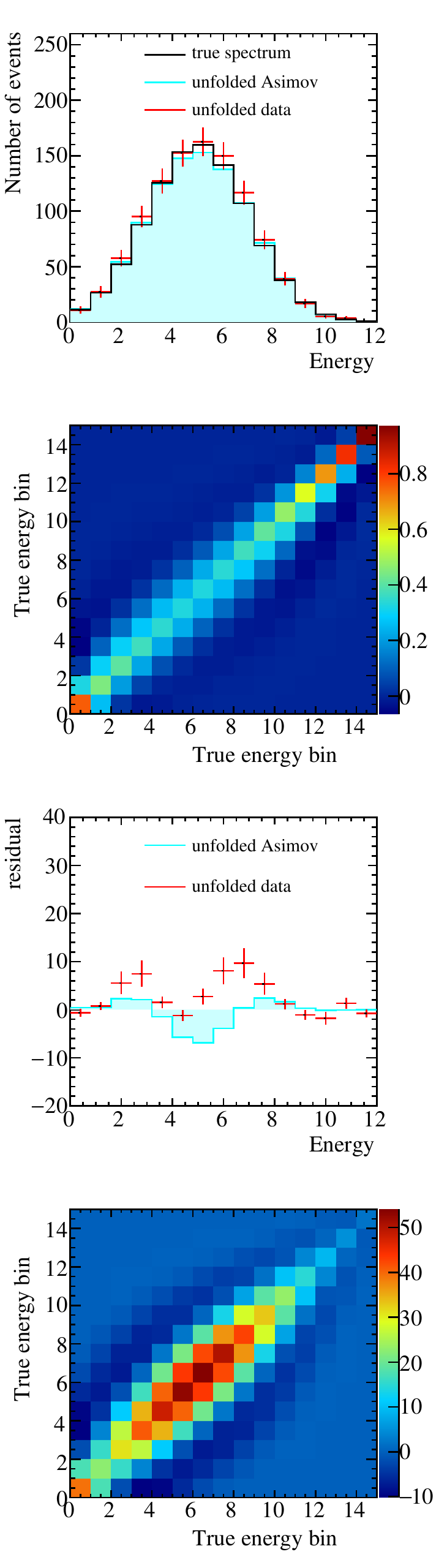}
\includegraphics[width=0.35\textwidth]{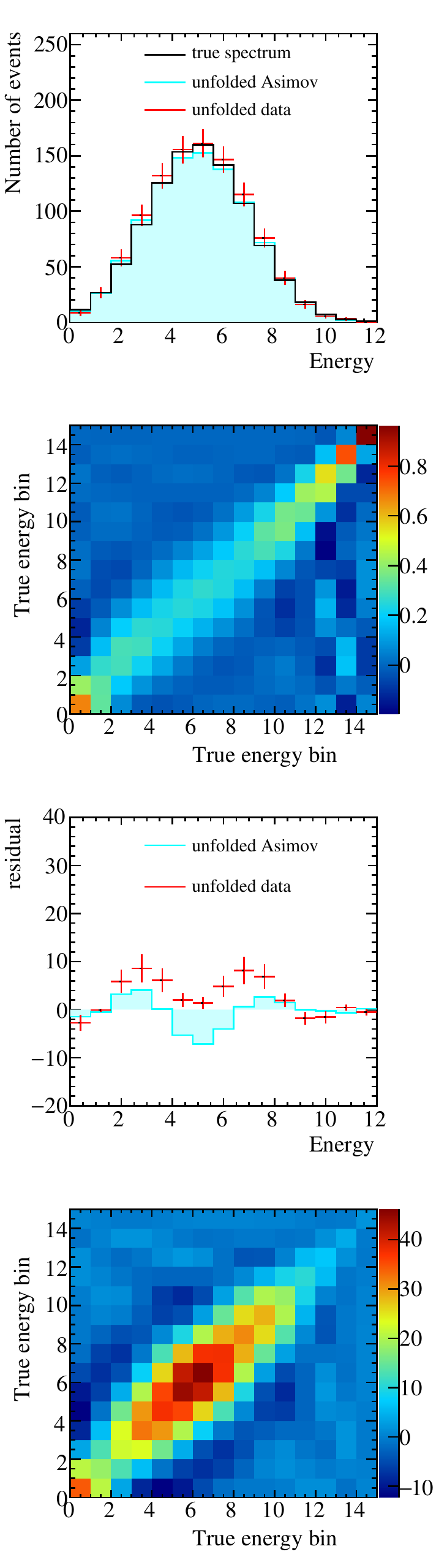}
  \end{center}
\caption{From top to bottom are: unfolded Asimov spectrum and data $\hat{s}$, additional smearing matrix $A_C$,
  residuals, and unfolded covariance matrix $Cov_{\hat{s}}$.
  The unfolded Asimov spectrum is obtained assuming the measurement is exactly same as the expectation
  based on true signal, i.e. without statistical fluctuation. 
  The unfolded data is obtained from the actual measured spectrum, i.e. with statistical fluctuation.
  In the case of unfolded spectrum, the residual is the same as bias.
  Left panels correspond to the regularization with $C_{2}$. Right panels correspond to
  Wiener-SVD with $C_{2}$. }
\label{fg:xs3}
\end{figure}

Figure~\ref{fg:xs3} shows the unfolded results based on the $C_2$ case for the regularization method (left panels) and the Wiener-SVD method (right panels). The unfolded spectra (top panel) and residual (the third panel to top)
are similar between the two methods. The additional smearing matrix (defined by Eq.~\ref{eq:additional}
and shown in the second panel to top) from regularization is more local than that of Wiener-SVD. This
is straightforward to understand, as the $C_2$ regularization constrains on
the smoothness and Wiener-SVD constrains on the signal to noise ratio in the effective frequency domain.
The bottom panels
of Fig.~\ref{fg:xs3} show the covariance matrices of the unfolded results. In comparison to
the diagonal covariance matrix of the measurement, the unfolded covariance matrix
is no longer diagonal due to the application of an additional smearing matrix.



\begin{figure} 
\includegraphics[width=0.5\textwidth, trim=10cm 0cm 0cm 0cm, clip=true]{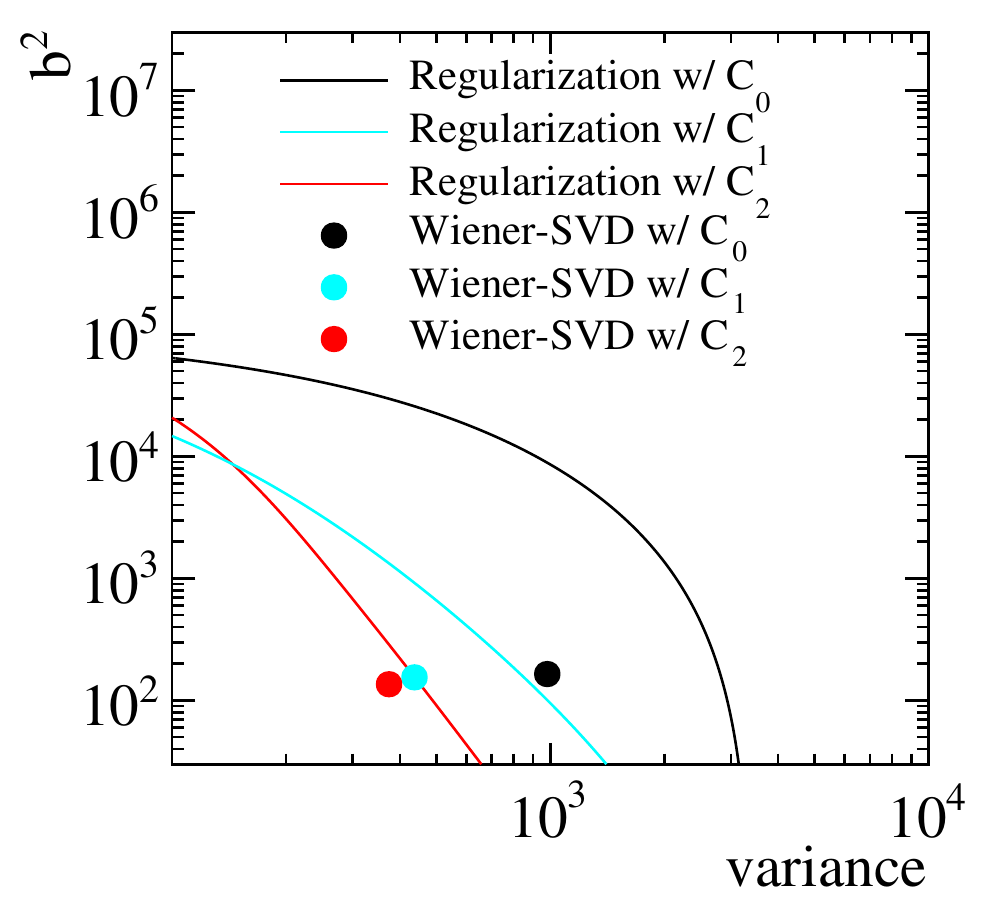}
\includegraphics[width=0.5\textwidth, trim=0cm 0cm 10cm 0cm, clip=true]{figs/xs_compare.pdf}
\caption{(Left) Bias squared $b^2$ v.s. variance $\sigma^2$ for various unfolding methods. (Right)
  MSE as a function of regularization strength $\tau$. The results from Wiener-SVD are shown as flat lines.
  See text for
more discussions.}
\label{fg:xs4}
\end{figure}

Figure~\ref{fg:xs4} shows the quantitative comparisons of the results from Tikhonov regularization
and Wiener-SVD unfolding in this example. In the left panel, the bias squared v.s. variance is plotted. For the Tikhonov regularization method, the regularization strength $\tau$ is scanned from 0 to 1.  As shown, at fixed variance (bias), the bias (variance)
of the Wiener-SVD result is smaller than those of the Tikhonov regularization method.
For both methods, the variance and bias of unfolded results with $C_2$ applied are better than those of $C_1$ and $C_0$. The right panel of Fig.~\ref{fg:xs4} shows the MSE as a
function of the regularization strength. We see that the MSEs of Wiener-SVD are smaller than the corresponding ones of Tikhonov regularization. 


\section{Data Unfolding Example: Reactor Neutrino Flux}\label{sec:reactor}


In the previous section, we illustrated the Wiener-SVD method with a low-statistics neutrino cross section extraction example, in which the problem is simplified by using a diagonal covariance matrix with only statistical uncertainties. 
In this section, we show a high-statistics example by constructing a toy reactor neutrino experiment to extract the reactor antineutrino
energy spectrum from the measured visible energy spectrum using the Wiener-SVD approach. In particular, we will implement a more realistic covariance matrix including both statistical and systematic uncertainties and illustrate how to construct the Wiener filter with a group of theoretically well-motivated models.  

In a typical reactor antineutrino experiment such as the Daya Bay experiment~\cite{An:2017osc}, the antineutrinos are detected through the inverse beta decay (IBD) process $\bar{\nu}_{e} + p \to e^{+} + n$. 
The positron gives a prompt signal including its kinetic energy and the two 511 keV annihilation gamma-rays, whereas the neutron after thermalization gets captured in the detector and yields a delayed signal.
Since the energy carried away by the recoil neutron is small, the neutrino energy $E_{\nu}$ can be approximately calculated from the prompt energy $E_{p}$ by $E_{p} \simeq  E_{\nu}$ - 0.8 MeV. 
The measurement of $E_{p}$ can be affected by a variety of systematic effects. For instance, in the Daya Bay experiment where the liquid scintillator is used as a calorimeter to determine the particle energy, the response from a particle's true energy to its visible energy is nonlinear. The nonlinearity is caused by both the quenching effect of the scintillator and the additional photons produced by the Cerenkov radiation. In addition, particles could lose energy in the non-scintillating materials, which further alters the visible energy. Various electronics nonlinear response can also occur and impact the total visible energy.   The resolution of $E_p$, typically $\sim$8\%, is mainly determined by the fluctuation of photoelectrons that follows the Poisson distribution. The gain variation, dark noise, and detector non-uniformity further add to the energy resolution. In order to construct the detector energy response matrix and the associated uncertainties, typically a comprehensive detector calibration campaign and data-Monte-Carlo comparison is necessary to fully understand these detector effects.

\begin{figure}[!h!tbp]
\includegraphics[width=\figwidth]{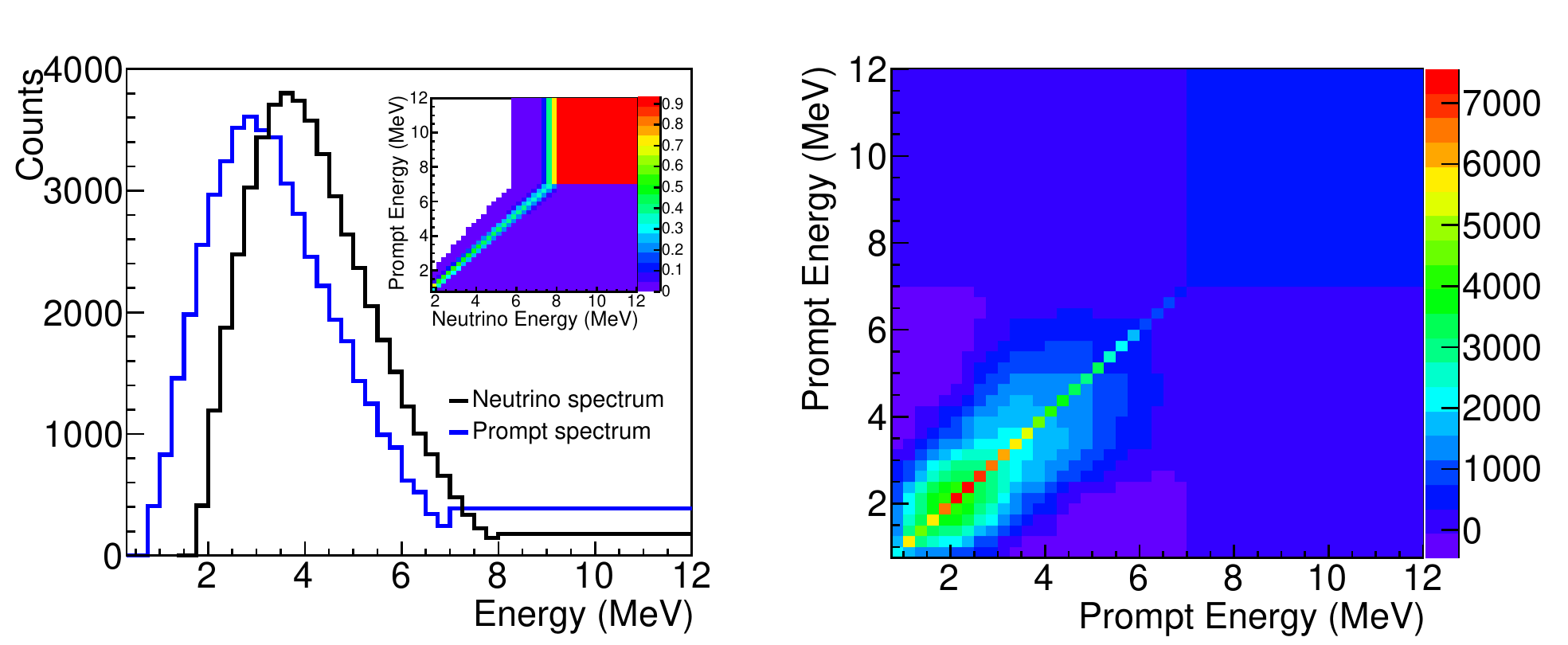}
\caption{(Left) {\em ``True''} neutrino spectrum $s_{true}$ (black) and the {\em ``measured''} prompt
  spectrum ${\bf m}$ (blue) used in this study. Each spectrum has 26 bins in total:
  the first 25 bins have the
  same bin width of 0.25 MeV and the last one has a larger binning of 4 (5) MeV for neutrino (prompt)
  spectrum. The inset panel gives the detector smearing matrix ${\bf r}$ used in this study.
   (Right) Covariance matrix $Cov$ for the prompt spectrum. The matrix is 26 $\times$ 26 and it
  has the same binning as the prompt spectrum in each dimension.}
\label{fig:reactor_input}
\end{figure}

We generated a $\sim$50k events toy reactor neutrino experiment using the Huber and Mueller reactor models\cite{Huber:2011flux, Mueller:2011flux} with a typical commercial reactor fission fractions for the four main isotopes: $^{235}$U, $^{238}$U, $^{239}$Pu, and $^{241}$Pu. We used the detector energy response reported by the Daya Bay experiment~\cite{An:2017reactor} to resemble a realistic situation. The covariance matrix used for the ``measured'' prompt spectrum in this toy study is generated based on the covariance matrix given in Ref.~\cite{An:2017reactor} with the corresponding statistics. Figure~\ref{fig:reactor_input} shows the inputs used for this toy study.
The left panel shows the {\em ``measured''} prompt spectrum (blue) that includes the detector
smearing effect and fluctuation due to uncertainties, the true neutrino spectrum (black), and the
detector smearing matrix (in the inset). The smearing matrix includes all detector response effects as mentioned above.  The right plot of Fig.~\ref{fig:reactor_input} shows the covariance
matrix for the prompt spectrum, which includes both the statistical and systematic uncertainties.

\begin{figure}[!h!tbp]
\includegraphics[width=\figwidth]{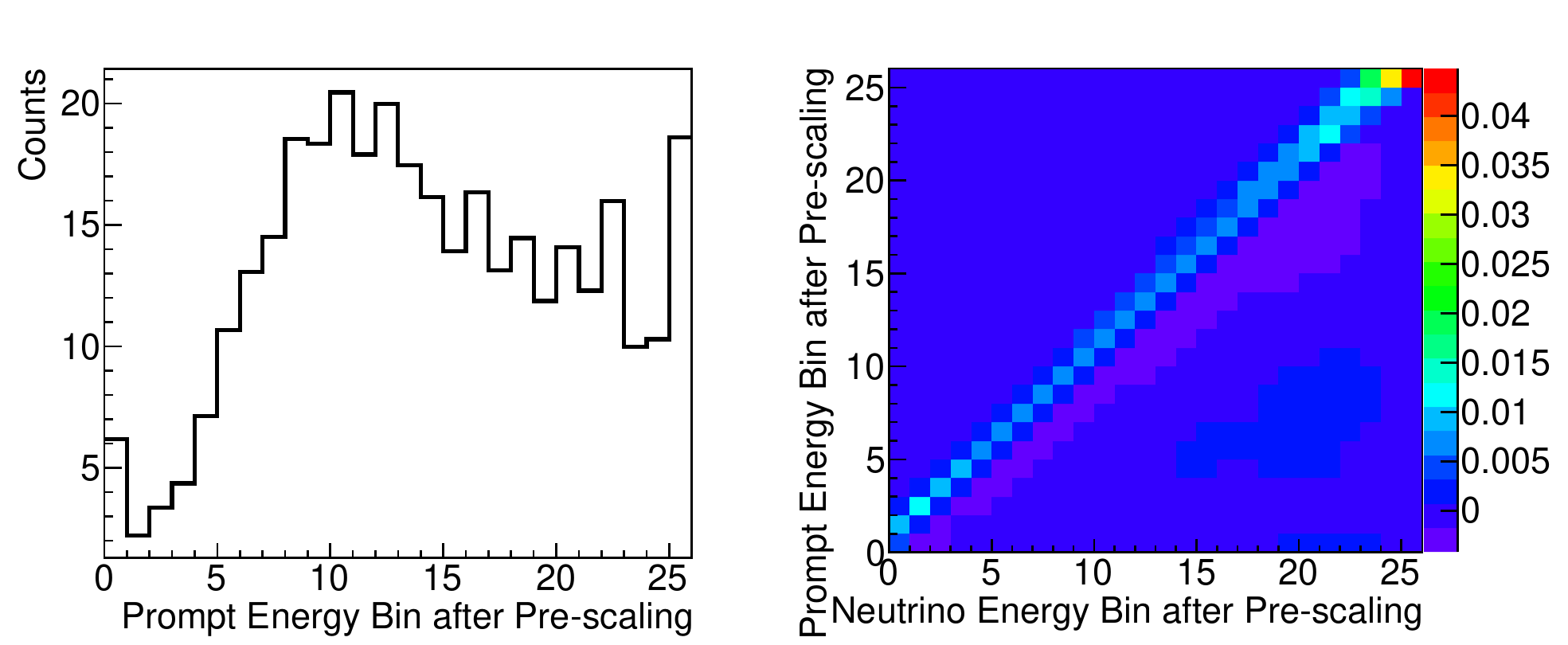}
\caption{(Left) The prompt spectrum after pre-scaling ($M:= Q\cdot {\bf m}$) is shown.
  (Right) The smearing matrix after pre-scaling ($R:=Q\cdot{} {\bf r}$) is shown.}
\label{fig:reactor_wsmear}
\end{figure}

As discussed in Sec.~\ref{sec:algorithm}, the first step of unfolding using SVD method is to do a
pre-scaling to normalize and remove correlations of uncertainties among bins.
Figure~\ref{fig:reactor_wsmear} shows the prompt spectrum (left plot) and smearing matrix (right plot)
after the pre-scaling. 
As can be seen,
both the pre-scaled prompt spectrum and the smearing matrix are quite different from the original
ones (i.e. Fig.~\ref{fig:reactor_input}).

In practice, the ``true'' model is always unknown, so it is not directly available for the purpose of constructing the Wiener
filter $W$. Instead, the Wiener filter can be constructed through a group of theoretical models using Eq.~\ref{eq:signal_exp1}. In this example, we consider a variety of
reactor flux models generated from the linear combinations of the calculations in Ref.~\cite{Dan:2015flux, Huber:2011flux, Mueller:2011flux}. The $\chi^{2}_{k}$ for model $s_{k}$ is then constructed by comparing the spectra of the prediction ${\bf m_{k}}$ $:=$ ${\bf r} \cdot s_{k}$ and the measurement ${\bf m}$. 
\begin{eqnarray}
  \chi^{2}_{k} &=& \left(  {\bf r} \cdot s_{k} - {\bf m} \right)^{T} \cdot Cov^{-1} \cdot \left( {\bf r}\cdot s_{k} - {\bf m} \right) \nonumber \\  
  &=& \left( {\bf m_{k}} - {\bf m} \right)^{T} \cdot Cov^{-1} \cdot \left( {\bf m_{k}} - {\bf m} \right) \nonumber \\
  &=& \left( M_{k} - M \right)^{2}
  \label{eq:reactor_chi}
\end{eqnarray}
where $M_{k}$ is ${\bf m_{k}}$ after pre-scaling: $M_{k}$ $:=$ $Q\cdot{\bf m_{k}}$.



\begin{figure}[!h!tbp]
\includegraphics[width=\figwidth]{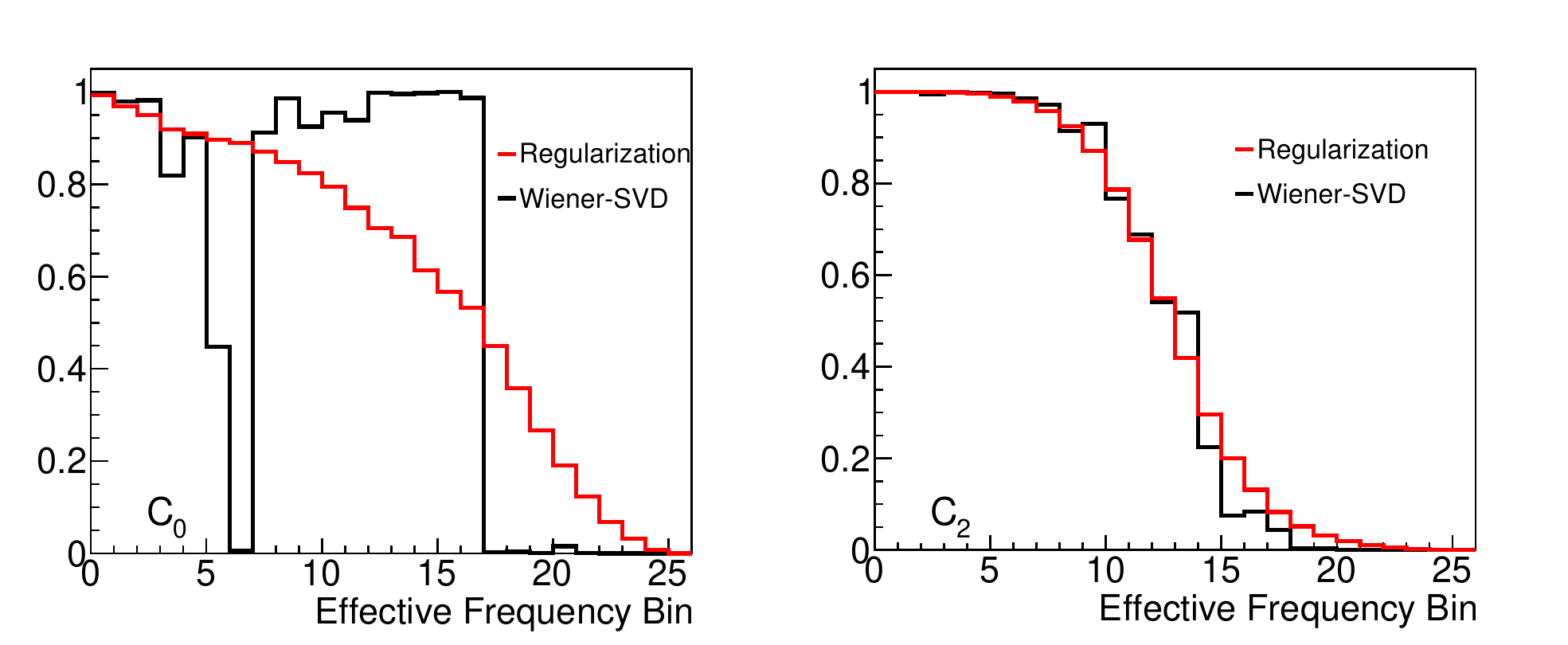}
\caption{Wiener filter $W$ and regularization filter $F$, which are constructed using different $C$ matrices,
  in their corresponding effective frequency domains. 
  The larger the bin number, the higher the frequency (i.e. the lower value the $d_{ii}$ is).
  (Left) Wiener filter (black) and regularization filter (red) constructed with $C_{0}$.
  (Right) Wiener filter and regularization filter constructed with $C_{2}$. }
\label{fig:reactor_wiener}
\end{figure}

Figure~\ref{fig:reactor_wiener} compares the Wiener filters $W$ and Tikhonov regularization filters $F$
in the effective frequency domain. They are constructed with the $C_{0}$ and $C_{2}$ matrices. 
For the regularization filters, the value of the regularization strength $\tau$ is chosen by
minimizing the MSE defined in Eq.~\ref{eq:reactor_mse1}.
As discussed previously, since the regularization filters only consider the $R$ in the effective
frequency domain, the larger the ``frequency'' the more the suppression. As can be seen
in both panels of Fig.~\ref{fig:reactor_wiener}, all of them behave as monotonically decreasing
functions. On the other hand, the Wiener-SVD method considers not only $R$ but also the signal to
noise ratio of each bin in the effective frequency domain. Therefore, the shapes of the Wiener filters
do not necessarily behave monotonically decreasing. For instance, The Wiener filter constructed with the $C_{0}$
(left panel of Fig.~\ref{fig:reactor_wiener}) has very large suppressions for the two medium frequency
bins (bin 6 and 7). Compared with the Wiener filter, the monotonicity of the regularization filter dictates
that it inevitably will keep more noise at these medium frequency bins and remove more signal
at some of higher frequency bins even when they are not very noisy. For filters constructed with
$C_2$ (right panel of Fig.~\ref{fig:reactor_wiener}), the Wiener filter has a similar shape as
the regularization filter. Nevertheless, the details of suppression at high frequency bins
are still different.



\begin{figure}[!h!tbp]
\includegraphics[width=\figwidth]{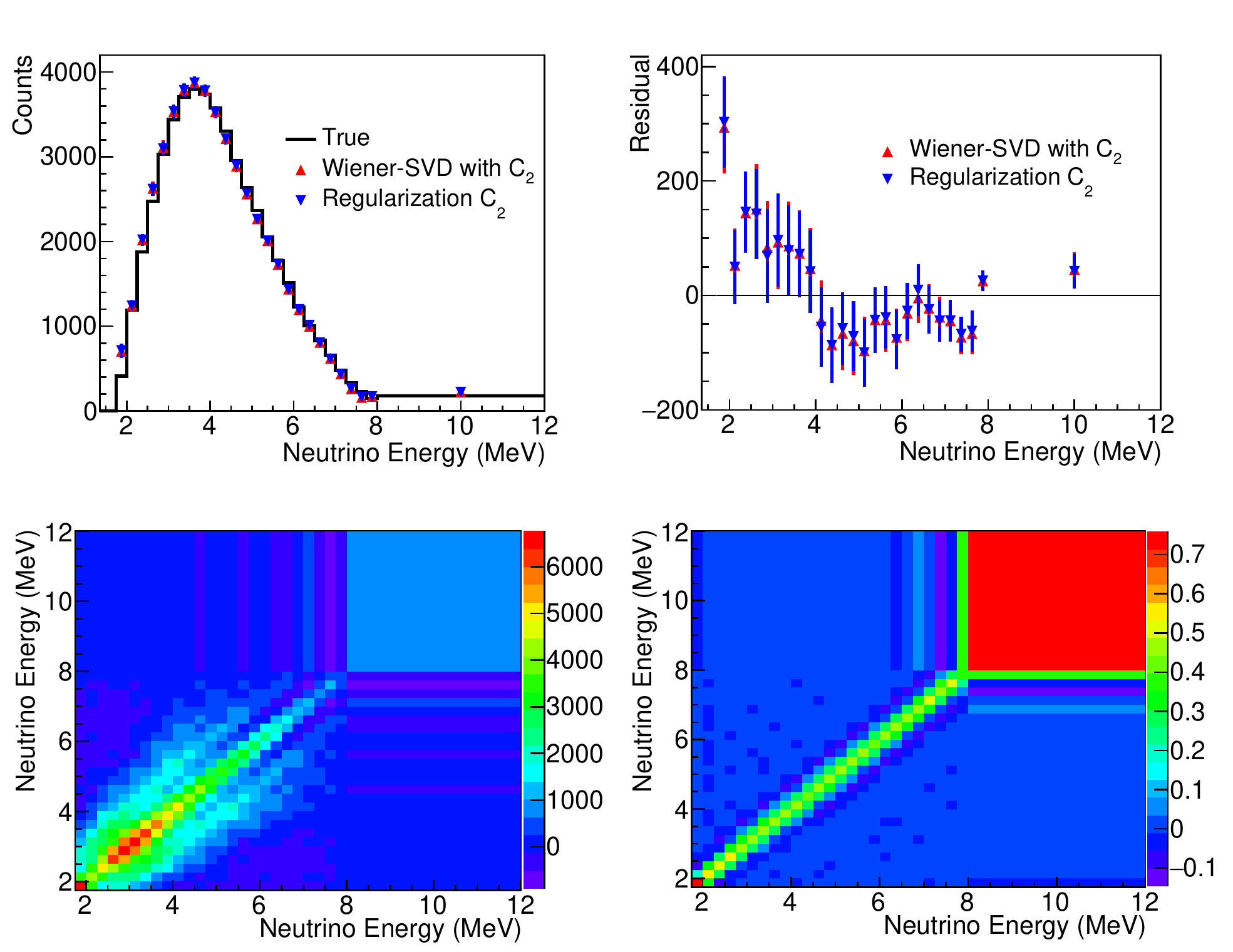}
\caption{(Top left) Comparison of the unfolded results $\hat{s}$ with true spectrum $s_{true}$.
  The red triangles are corresponding to the Wiener-SVD unfolding with the $C_{2}$ matrix.
  The blue triangles represent the unfolded result using regularization method with the $C_{2}$ matrix
  and regularization strength $\tau$ at 2.4$\times$10$^{-5}$. The error bars shown in the plot
  are taken from the square root of the diagonal elements of the unfolded covariance matrix
  $Cov_{\hat{s}}$ for each method. (Top right) Residual of unfolded spectrum with true spectrum (Bottom left) The covariance matrix for the unfolded spectrum
  $Cov_{\hat{s}}$ using Wiener-SVD method. (Bottom right) The additional smearing
  matrix $A_{c}$ for the Wiener-SVD unfolding method.}
\label{fig:reactor_unfold}
\end{figure}

\begin{figure}[!h!tbp]
\includegraphics[width=\figwidth]{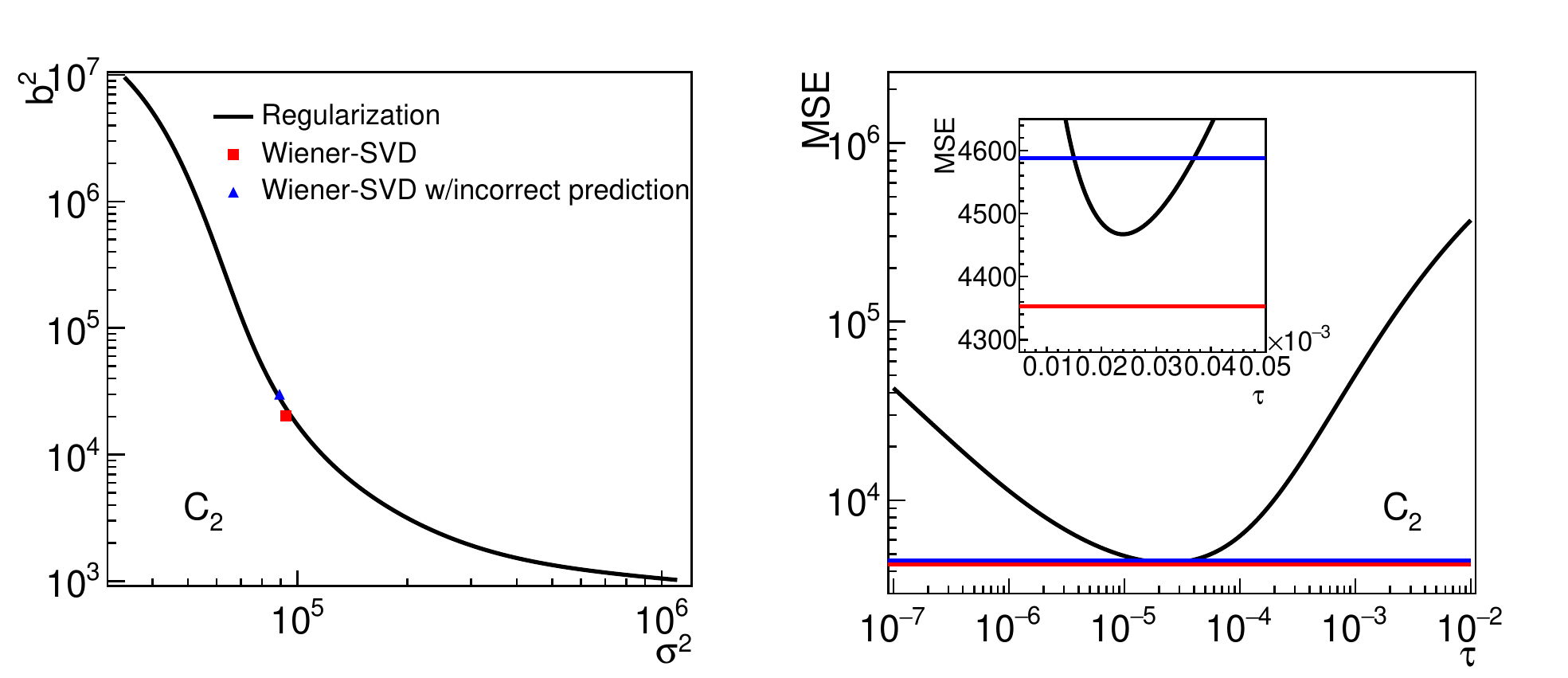}
\caption{(Left) The variance $\sigma^2$ versus bias square $b^2$ plot of the unfolded results $\hat{s}$
  for the Wiener-SVD and regularization methods. Both methods using the $C_{2}$ matrix. The red square is the
  result of the Wiener filter that is constructed from the a large number of predictions based on the models
  in Ref.~\cite{Dan:2015flux}. The blue triangle is the result of the Wiener filter that is constructed
  using one model that is quite different from the true model. The black curve is a scan of a wide range
  of $\tau$ for the regularization method. (Right) MSE values vs. $\tau$ for regularization method are shown.
  Wiener-SVD results are shown as the red and blue flat lines. The inset shows the zoom in of the valley
  in the plot. The Wiener-SVD unfolded result (red) produces a smaller MSE than any unfolded result from
  regularization method. This is no longer true when the Wiener filter is constructed using an improper
  expectation (blue) and the regularization result would also be altered which is not shown in this figure.}
\label{fig:reactor_mse}
\end{figure}

The unfolded results of the Wiener-SVD and regularization methods can be seen in Fig.~\ref{fig:reactor_unfold}.  
The regularization unfolding uses the regularization strength $\tau$ = 2.4$\times$10$^{-5}$ through minimizing the MSE in Eq.~\ref{eq:reactor_mse1}. For simplicity, only results with the using of a $C_2$ matrix are shown. Both methods produce reasonable unfolded results. To compare the Wiener-SVD and
regularization unfolded results with different values of $\tau$, we plot the total variance
versus total bias square in the left panel of Fig.~\ref{fig:reactor_mse}. The bias is calculated using Eq.~\ref{eq:bias} with $\overline{s}$ set to the model that has the smallest $\chi^{2}$ value (see Eq.~\ref{eq:reactor_chi}). The black curve
is from the regularization method with a wide range of $\tau$. The red square is from the
Wiener-SVD method described previously. Similar to the cross section example in the previous section, 
at the same variance (bias), the Wiener-SVD method has a smaller bias (variance). The right panel of
Fig.~\ref{fig:reactor_mse} shows the corresponding MSE values. Again, similar to the previous section,
the Wiener-SVD unfolded result has a smaller MSE than any unfolded result from the regularization
method. 
To illustrate the necessity of using $\bar{s}$ to represent the unknown $s_{true}$, we also show an unfolded result (blue triangle)
with a Wiener filter constructed using an improper expectation (quite different from the $\bar{s}$).
In this case, the MSE of the unfolded results with the Wiener filter is no longer the smallest.
In practice, it is crucial to use Eq.~\ref{eq:signal_exp1} to evaluate the signal expectation in
order to achieve the optimal bias and variance in the Wiener-SVD approach.
This is the common issue for all kinds of unfolding approaches, and in fact the best result of regularization method (minimum point in Fig.~\ref{fig:reactor_mse} right panel) would also be altered by comparing the unfolded result with the improper signal expectation.

\section{Discussions and Recommendations}\label{sec:discussion}

Based on the examples in previous two sections, we make the following
recommendations regarding data unfolding:
\begin{itemize}
\item The SVD-based unfolding methods (traditional regularization filter or Wiener filter)
  are equivalent to replacing the detector smearing matrix with a new smearing matrix $A_C$.
  The application of this new smearing matrix is crucial to suppress the large oscillation
  (high variance) of the direct matrix inversion unfolded results.
  In evaluating bias, the $A_C$ is applied to the true spectrum, which leads to bias.
\item We recommend to report this new smearing matrix $A_C$ in the publication
  together with the unfolded results to
  enable a more direct comparison of expectations (e.g. from new theoretical calculations)
  with the unfolded results. In practice, the new smearing matrix should be applied to
  the theoretical calculation before comparing to the unfolded results. Through reporting
  the new smearing matrix $A_C$, one can avoid including the bias due to unfolding in
  the final uncertainties.
\item For SVD-based approach, the $C_2$ typically yields a better result than those of
  $C_1$ and $C_0$. 
\item The covariance matrix, variance, and the bias of the unfolded results can be calculated
  with Eq.~\ref{eq:unfold_cov}, Eq.~\ref{eq:var}, and Eq.~\ref{eq:bias}, respectively.
\item The Wiener filter should be constructed by Eq.~\ref{eq:signal_exp1}, which takes
  into account the measurement given a range of prior expectations.
\end{itemize}

From these two toy examples shown in Sec.~\ref{sec:xs} and Sec.~\ref{sec:reactor}, we can
conclude the following pros and potential cons for the Wiener-SVD approach in comparison to the
Tikhonov regularization approach:
\begin{itemize}
\item The Wiener-SVD approach is free from a regularization parameter, which is
  required to be optimized for the regularization approach. This is achieved by utilizing the
  expectations of signal and noise, which can be viewed as a direct determination of an effective
  regularization parameter. 
  At a fixed variance (bias), unfolded results from the Wiener-SVD method gives a better bias
  (variance) than the regularization method. This presumably is due to the optimized
  signal to noise ratio in the effective frequency domain for the Wiener filter.
  When evaluated with the MSE (metric defined in Eq.~\ref{eq:reactor_mse1}), the unfolded results
  based on the Wiener filter is comparable and sometimes better than
  the best result from the regularization approach. 
\item For $C_1$ and $C_2$, the traditional regularization method pulls the estimator
  towards smoothness, even though the true distribution is not necessarily so. The Wiener
  filter considers the signal to noise ratio in the effective frequency domain, which does
  not require smoothness and is more general.
\item To construct the Wiener filter, an estimation of the true model is required.
  The unfolded results in the Wiener-SVD method is strictly model dependent. Such dependence
  is reduced when the estimation of the true model takes into account the actual measurement
  as illustrated in Eq.~\ref{eq:signal_exp1}. In addition, the different choices of the true
  model only affects the construction of the new smearing matrix. By reporting the new
  smearing matrix, the model dependence of the results can be avoided, as the new smearing
  matrix can be applied to the other expectations to be tested.
\item As shown in Fig.~\ref{fg:xs3}, the new smearing matrix of the Wiener-SVD is less
  localized than
  those of the regularization method. This could be a potential disadvantage of the
  Wiener-SVD approach, but can again be mitigated by reporting the new smearing matrix.
\end{itemize}

The derivation of the Wiener filter construction as shown in Eq.~\ref{eq:wiener_reg1} is
based on the covariance matrix and SVD decomposition of the smearing matrix $R \cdot C^{-1}$.
In the case when the uncertainties cannot be simply expressed via a covariance matrix
(i.e. Gaussian approximation), the Wiener filter can still be constructed if the
smearing matrix can be constructed. In this case,
the expectation of signal square can still be constructed using Eq.~\ref{eq:signal_exp1}.
The expectation of noise square would be obtained through a Monte Carlo approach. 

\section{Summary}\label{sec:summary}

Inspired by the deconvolution technique employed in the digital signal processing, we introduce a new
unfolding technique based on the Wiener filter and SVD technique for HEP data analysis. 
Through maximizing the signal to noise ratios in the effective frequency domain, the
Wiener-SVD unfolding avoids the scanning of any regularization parameter in the
traditional approaches. Through a couple examples, we show that the unfolded
results from the Wiener-SVD method generally have a smaller bias (variance) at fixed
variance (bias) than the unfolded results from the Tikhonov regularization method.
The overall MSE averaging the total bias and variance is also generally smaller
for the Wiener-SVD method. These features support the Wiener-SVD method as an
attractive option for the data unfolding problem. An implementation of the Wiener-SVD
method can be found in Ref.~\cite{code}.


\acknowledgments

We thank Tom Junk, Clark McGrew for fruitful discussions and David Caratelli for carefully
reading of the manuscript. This work is supported by the U.S. Department of Energy,
Office of Science, Office of High Energy Physics, and Early Career
Research Program under contract number DE-SC0012704.




\bibliographystyle{hunsrt}

\bibliography{unfolding}{}

\end{document}